\documentclass[useAMS,usenatbib,usegraphicx,lscape,rotating]{mn2e}
\newcommand\HI{\mbox{H\thinspace{\sc i}}}
\newcommand\HII{\mbox{H\thinspace{\sc ii}}}
\newcommand\kms{\mbox{km~s$^{-1}$}}
\newcommand\nodata{ ~$\cdots$~}

\title{A multi-resolution analysis of the radio-FIR correlation in the
  Large Magellanic Cloud} \author[Hughes et
  al.]{A. Hughes$^{1,2}$\thanks{Email: ahughes@astro.swin.edu.au},
  T. Wong$^{2,3}$\thanks{ARC-CSIRO Postdoctoral Linkage Fellow},
  R. Ekers$^{2}$, L. Staveley-Smith$^{2}$, \newauthor
  M. Filipovic$^{2,4}$, S. Maddison$^{1}$, Y. Fukui$^{5}$,
  N. Mizuno$^{5}$ \\ $^1$ Centre for Supercomputing and Astrophysics,
  Swinburne University of Technology, Hawthorn VIC 3122, Australia
  \\ $^2$ CSIRO Australia Telescope National Facility, PO Box 76,
  Epping NSW 1710, Australia \\ $^3$ Department of Astrophysics and
  Optics, School of Physics, University of New South Wales, Sydney NSW
  2052, Australia \\ $^4$ University of Western Sydney, Locked Bag
  1797, Penrith South, DC, NSW 1797, Australia \\ $^5$ Department of
  Astrophysics, Nagoya University, Chikusa-ku, Nagoya 464-8602,
  Japan\\ }
\begin{document}

\date{Accepted 2006 April 20. Received 2006 April 19; in original form
  2006 January 27}

\maketitle

\label{firstpage}

\begin{abstract}
\label{sect: abstract}
 We investigate the local correlation betwen the 1.4~GHz radio
 continuum and 60~$\mu$m far-infrared (FIR) emission within the Large
 Magellanic Cloud (LMC) on spatial scales between 0.05 and 1.5~kpc. On
 scales below $\sim$1~kpc, the radio-FIR correlation is clearly better
 than the correlation of the cold gas tracers with either the radio or
 the FIR emission. For the LMC as a whole, there is a tight
 correlation between the radio and FIR emission on spatial scales
 above $\sim$50~pc. By decomposing the radio emission into thermal and
 non-thermal components, however, we show that the scale on which the
 radio-FIR correlation breaks down is inversely proportional to the
 thermal fraction of the radio emission: regions that show a strong
 correlation to very small scales are the same regions where the
 thermal fraction of the radio emission is high. Contrary to previous
 studies of the local radio-FIR correlation in the LMC, we show that
 the slope of the relation between the radio and FIR emission is
 non-linear. In bright star-forming regions, the radio emission
 increases faster than linearly with respect to the FIR emission
 (power-law slope of $\sim1.2$), whereas a flatter slope of
 $\sim0.6-0.9$ applies more generally across the LMC. Our results are
 consistent with a scenario in which the UV photons and cosmic rays in
 the LMC have a common origin in massive star formation, but the
 cosmic rays are able to diffuse away from their production sites. Our
 results do not provide direct evidence for coupling between the
 magnetic field and the local gas density, but we note that
 synchrotron emission may not be a good tracer of the magnetic field
 if cosmic rays can readily escape the LMC.
\end{abstract}


%%%%%%%%%%%%%%%%%%%%%%%%%%%%%%%%
%%%%%%%%%%%%%%%%%%%%%%%%%%%%%%%%
\section{Introduction}
%%%%%%%%%%%%%%%%%%%%%%%%%%%%%%%%
%%%%%%%%%%%%%%%%%%%%%%%%%%%%%%%%
\label{sect:intro}

\begin{figure*}
\includegraphics[width=170mm]{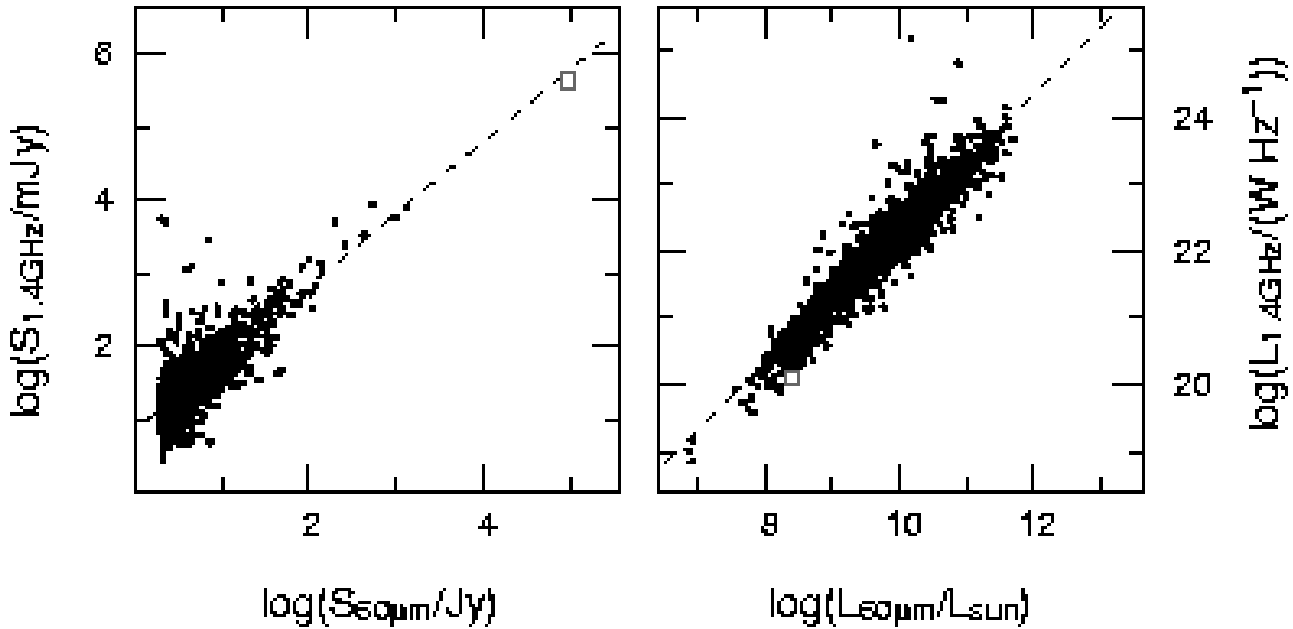}
\caption{Plot of 1.4~GHz vs IRAS 60~$\mu$m flux density ({\it left}), and
  1.4~GHz radio luminosity vs. IRAS 60~$\mu$m luminosity ({\it right}) for
  the \citet{yunetal01} sample of 1809 galaxies. In both panels, the
  dashed line is a linear least squares fit to the data. The
  LMC is indicated as an open square. }
\label{fig:yunplot}
\end{figure*}

The tight, almost ubiquitous correlation between the far-infrared
(FIR) and radio continuum emission in star-forming galaxies remains
one of the most robust, yet puzzling, relationships in extragalactic
astronomy \citep[see e.g.][see also Fig.~\ref{fig:yunplot}]{yunetal01}. The
correlation is essentially linear over five orders of magnitudes in
luminosity with a dispersion of less than 50\%. It applies to galaxies
beyond $z=1$ \citep{appletonetal04} and encompasses a diverse range of
galaxy types, including normal barred and unbarred spirals, dwarf and
irregular galaxies, starbursts, Seyferts and radio-quiet quasars
\citep[for a review, see][]{condon92}.\\
 
The conventional explanation for the radio-FIR correlation invokes
massive star formation. In this model, the FIR emission is mostly
thermal emission from dust that has been heated by the ultra-violet
(UV) radiation from young massive stars.  These same stars power the
thermal radio emission of star-forming regions, and rapidly evolve to
the supernovae whose remnants are responsible for accelerating the
cosmic-ray electrons that produce the non-thermal component of the
radio emission.  The radio emission of normal spiral galaxies at
centimetre wavelengths is predominantly non-thermal \citep{condon92},
while the total FIR emission comprises both a warm dust component,
heated in \HII\ regions by massive ($>$20 $\rm{M}_{\odot}$) ionising
stars, and a significant cool dust component that is heated by the
interstellar radiation field \citep[e.g.][]{walterbosschwering87}.
Efforts to decompose the radio emission into thermal and non-thermal
components, and the FIR emission into warm and cool dust components,
indicate that the overall radio-FIR correlation is constituted by a
strong correlation between the thermal radio and the warm dust
emission, and a weaker correlation between the non-thermal radio and
the cool dust emission \citep{hoernesetal98}.\\

Although a correlation between warm dust and thermal radio emission
from ionised gas is hardly surprising, the mechanism by which the dust
grains couple to synchrotron emission arising through the interactions
of relativistic electrons with the ambient magnetic field remains
highly uncertain.  \citet{heloubicay93} have suggested a coupling
between magnetic field strength and gas density, which would ensure a
tight radio-FIR correlation in spite of the sensitive dependence of
the sychrotron emission on the magnetic field.  Variations on this
idea have been discussed by \citet{hoernesetal98} and
\citet{murgiaetal05}, the latter also connecting the radio-FIR
correlation to the observed CO-radio correlation, as CO traces the
dense molecular gas.  An important constraint on these models is the
scale on which the correlation breaks down, as this would shed light
on the physical mechanism of the coupling (e.g., static pressure
equilibrium predicts that the correlation should hold on scales
greater than the scale height of a galaxy's gas disc).\\

Observationally, FIR and radio maps of nearby spiral galaxies
generally show good correspondence, although studies of nearby spirals
are limited to scales above a few hundred parsecs
\citep[e.g.][]{hoernesetal98,hippeleinetal03}. On much smaller scales,
detailed maps of the gas within several kiloparsecs of the Sun
indicate that the radio emission from star-forming regions is
predominantly thermal, and does not follow the radio-FIR correlation
for integrated galaxy luminosities
\citep[e.g.][]{boulangerperault88,haslamosborne87}. Galactic supernova
remnants do not seem to be well-correlated with FIR or mid-infrared
(MIR) emission, or with sites of star formation
\citep{cohengreen01,whiteoakgreen96}. A tight, linear correlation
between the integrated radio and FIR emission in galaxies is thus not
at all expected from observations of the local interstellar medium
(ISM). In light of this conflict, and the different effective scales
of proposed mechanisms for the origin of the radio-FIR correlation, it
seems timely to study the radio and FIR emission of a galaxy on
intermediate spatial scales between those accessible in the Milky Way
and in nearby spirals.\\

The aim of this study is to investigate the radio-FIR correlation
within the Large Magellanic Cloud (LMC).  The LMC is a gas-rich dwarf
galaxy that exhibits clear signs of active star formation. Our global
view of the LMC is complemented by sufficient spatial resolution to
study the ISM structure in detail over a wide range of wavebands.  At
an assumed distance of 50.1~kpc \citep{alves04}, 1 arcmin corresponds
to 14.5~pc. The proximity of the LMC thus allows a detailed comparison
of the radio and FIR emission on scales down to 20~pc (for the highest
resolution radio and FIR datasets that are currently available).\\

In addition to standard pixel-by-pixel techniques, we use a wavelet
cross-correlation method for our analysis that allows us to quantify
the correlation between different gas and dust tracers on spatial
scales between $\sim$0.04 and 2.0~kpc. To date,
wavelets have enjoyed a limited (though rapidly increasing)
application within astrophysics, but their ability to isolate signal
simultaneously in the Fourier and spatial domains makes them ideal for
analysing scale-dependent correlations in astronomical images.\\

The structure of this paper is as follows. In Section \ref{sect:data}
we describe our data sets and reduction procedures.  Our analysis
methods and results are presented in Section \ref{sect:analysis}.  We
discuss our results in the context of current models for the origin of
the radio-FIR correlation in Section \ref{sect:discussion}.  Our
conclusions are presented in Section \ref{sect:conclusions}.

%%%%%%%%%%%%%%%%%%%%%%%%%%%%%%%%%%%%%%
%%%%%%%%%%%%%%%%%%%%%%%%%%%%%%%%%%%%%%
% FIGURE 1
%%%%%%%%%%%%%%%%%%%%%%%%%%%%%%%%%%%%%%
%%%%%%%%%%%%%%%%%%%%%%%%%%%%%%%%%%%%%%

%%%%%%%%%%%%%%%%%%%%%%%%%%%%%%%%
%%%%%%%%%%%%%%%%%%%%%%%%%%%%%%%%
\section{Observational Data}
%%%%%%%%%%%%%%%%%%%%%%%%%%%%%%%%
%%%%%%%%%%%%%%%%%%%%%%%%%%%%%%%%
\label{sect:data}

\begin{figure*}
\includegraphics[width=15cm]{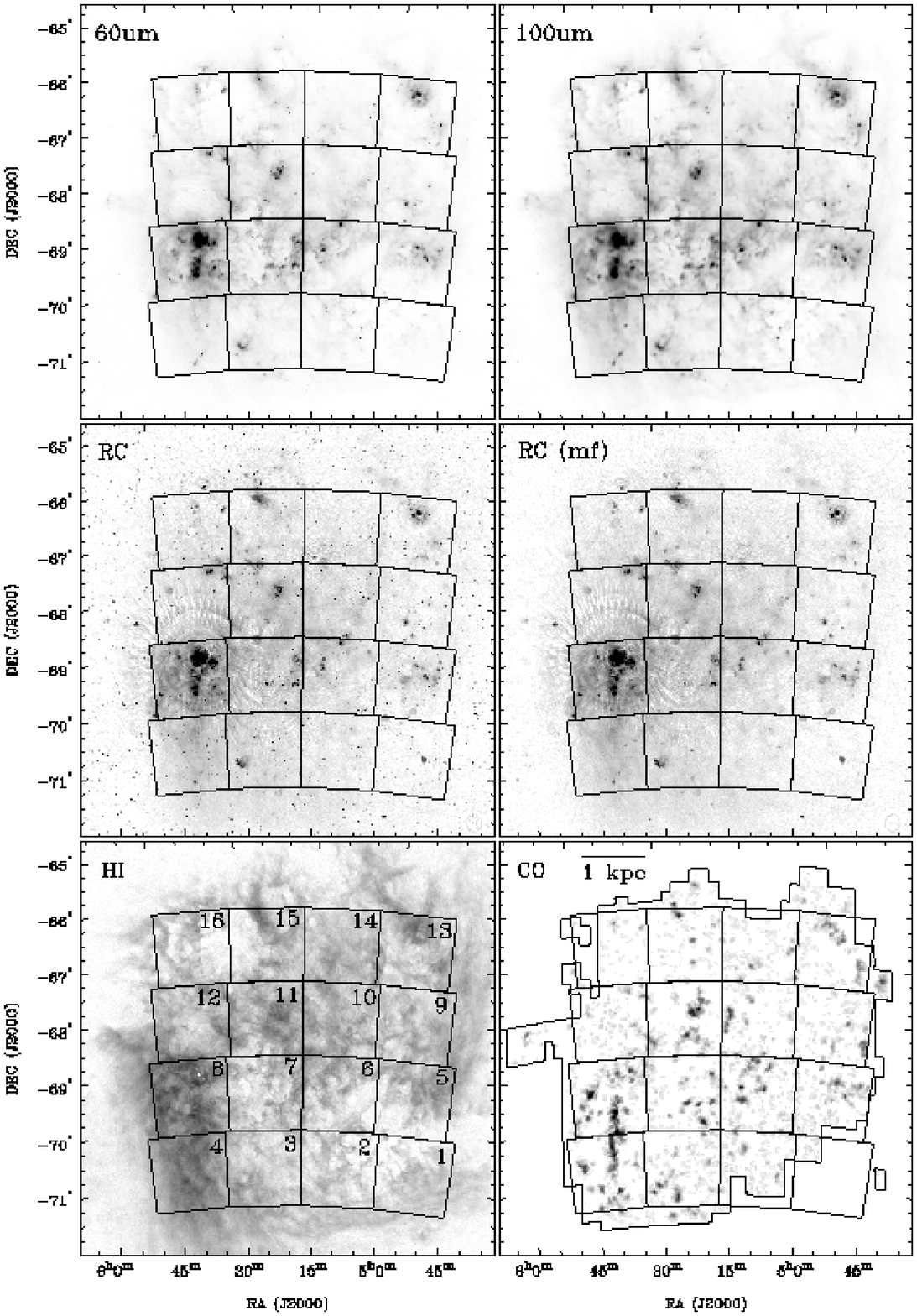}
\caption{Maps of the Large Magellanic Cloud at different wavelengths,
  overlaid by the sub-region fields described in the text. The images
  correspond to HIRAS 60~$\mu$m emission ({\it top left}), HIRAS
  100~$\mu$m emission ({\it top right}), ATCA+Parkes 1.4~GHz radio
  continuum ({\it middle left}), median filtered ATCA+Parkes 1.4~GHz
  radio continuum ({\it middle right}), ATCA+Parkes \HI\ integrated
  intensity ({\it bottom left}), and NANTEN CO J=1--0 integrated
  intensity ({\it bottom right}).  A square-root intensity scale has
  been used in displaying all images.}
\label{fig:LMCmap}
\end{figure*}

The images we have used for this study are described in detail below,
and summarized in Table~\ref{tbl:images}.  Originating from a variety
of instruments, they cover different fields of view at different
angular resolutions and pixel sizes.  In order to compare the data,
all maps were regridded to a common pixel size of 20 arcsec and a
common J2000 map centre of $\alpha$=5$^{\rm h}$20$^{\rm m}$38\fs6,
$\delta$=$-68$\degr41\arcmin22\arcsec.  Images of 1350$^2$ pixels,
covering a roughly 7\fdg5 $\times$ 7\fdg5 region, were produced using
an NCP projection, which is essentially an orthographic (SIN)
projection with a tangent point at the north celestial pole.  The
final regridded, truncated images are shown in Fig.~\ref{fig:LMCmap}.
Note that the region thus defined, which we refer to as the ``whole
LMC,'' is completely covered by all maps except for the CO map, which
has a somewhat irregular sky coverage (Fig.~\ref{fig:LMCmap}).  Total
fluxes and flux densities measured within the selected region are
given in Table~\ref{tbl:images}.\\

Since we are interested in spatial variation of the radio-FIR
correlation across the LMC, the maps were divided into 16 sub-regions,
as shown in Fig.~\ref{fig:LMCmap}. Each sub-region is 1\fdg35 $\times$
1\fdg35 in size, with 256 $\times$ 256 pixels and a pixel scale of 19
arcsec. The centre positions of the 16 sub-region maps are listed in
Table~\ref{tbl:SRpos}, along with brief notes about the individual
fields.  Note the wide variation in flux density among sub-regions in
the 1.4~GHz and 60~$\mu$m maps: for instance, sub-region 8, which
covers the 30 Doradus region, contains roughly a third of the total
radio and FIR emission in the LMC.\\

%%%%%%%%%%%%%%%%%%%%%%%%%%%%%%%%
%%%%%%%%%%%%%%%%%%%%%%%%%%%%%%%%
%% TABLE 1, FIGURE 1          %%
%%%%%%%%%%%%%%%%%%%%%%%%%%%%%%%%
%%%%%%%%%%%%%%%%%%%%%%%%%%%%%%%%

\begin{table}
  \caption{Properties of the datasets used in this study, prior to any
    convolution or regridding.  The fluxes and flux densities are
    measured over a roughly 7\fdg5 $\times$ 7\fdg5 region covering the
    main body of the LMC.}
  \label{tbl:images} 
    \begin{tabular}{@{}lllll@{}}
   \hline
   Band & Telescope(s) & Resolution & Pixel scale & Flux Density\\
        &              & (arcmin)   & (arcmin)    & (Jy)\\
   \hline
   60\,$\mu$m & IRAS (HIRAS) & 1.3 & 0.25 & $9.64 \times 10^4$\\
   60\,$\mu$m & IRAS (IRIS) & 4.0 & 1.5 & $9.61 \times 10^4$\\
   100\,$\mu$m & IRAS (HIRAS) & 3.0 & 0.25 & $2.09 \times 10^5$\\
   100\,$\mu$m & IRAS (IRIS) & 4.3 & 1.5 & $2.04 \times 10^5$\\
   6.3\,cm & Parkes & 4.8 & 1.8 & $2.96 \times 10^2$\\
   21\,cm & Parkes & 15.2 & 5.0 & $4.71 \times 10^2$\\
   21\,cm & ATCA+Parkes & 0.67 & 0.22 & $4.26 \times 10^2$\\
  \hline
   Band & Telescope(s) & Resolution & Pixel scale & Flux\\
        &              & (arcmin)   & (arcmin)    & (Jy \kms)\\  
   \hline
   \HI & ATCA+Parkes & 1.0 & 0.33 & $3.98 \times 10^5$\\
   CO & NANTEN & 2.6 & 2.0 & $8.80 \times 10^5$\\
   \hline
    \end{tabular}
\end{table}

%%%%%%%%%%%%%%%%%%%%%%%%%%%%%%%%
\subsection{Far-Infrared Data}
%%%%%%%%%%%%%%%%%%%%%%%%%%%%%%%%
\label{sect:IRASdata}

Infrared Astronomical Satellite (IRAS) mosaics of the LMC in the
60~$\mu$m and 100~$\mu$m bands were kindly provided to us by M. Braun
and have been discussed by \citet{braunetal98}.  These data had been
processed using the HIRAS algorithm, which uses the Pyramid Maximum
Entropy (PME) method to improve upon the nominal resolution of the
original IRAS survey data (4.0$\pm$0.2 arcmin for the 60~$\mu$m band,
and 4.3$\pm$0.2 for the 100~$\mu$m band) \citep{bontekoeetal94}.  The
PME algorithm relies on the local sampling pattern in the IRAS survey,
so the final resolution of HIRAS images varies spatially across a map.
By fitting Gaussian profiles to point-like sources, we found that the
resolution of the HIRAS maps was $\sim$1.3 arcmin for the 60~$\mu$m
data and $\sim$3.0 arcmin for the 100~$\mu$m data.  Measuring the
average brightness for regions of sky away from the LMC, we determined
a background level of 1.25 MJy sr$^{-1}$ for the 60~$\mu$m map and 3.0
MJy sr$^{-1}$ for the 100~$\mu$m map.  These constant values were
subtracted from the maps.  A complete description of IRAS data
products can be found on the IRAS Documentation
website.\footnote{http://irsa.ipac.caltech.edu/IRASdocs/toc.html} \\

To verify the flux density calibration of the HIRAS data, IRIS
60~$\mu$m and 100~$\mu$m maps were obtained from
Dr.\ Miville-Desch{\^e}nes at CITA
\citep{mivilledescheneslagache05}. The IRIS images are at the same
resolution as the original IRAS survey data, but have better zodiacal
light subtraction and destriping than the equivalent IRAS Sky Survey
images.  The flux density calibration of IRIS is consistent with DIRBE data,
but we found it was necessary to apply a small zero-level correction
to the maps. Measuring the average brightness for regions of sky away
from the LMC, we determined a background level of 0.85 MJy sr$^{-1}$
for the 60~$\mu$m map and 3.0 MJy sr$^{-1}$ for the 100~$\mu$m map. We
subtracted these constant values from the maps. It is worth noting
that the background levels measured for the IRIS maps are very similar
to the background levels measured for the HIRAS maps, which increases
our confidence in the flux density calibration of the HIRAS maps. A complete
description of IRIS data products can be found on the IRIS
website.\footnote{http://www.cita.utoronto.ca/$\sim$mamd/IRIS/IrisOverview.html}\\

%%%%%%%%%%%%%%%%%%%%%%%%%%%%%%%%%%%%%%%%
%%%%%%%%%%%%%%%%%%%%%%%%%%%%%%%%%%%%%%%%
%
% TABLE 2
%
%%%%%%%%%%%%%%%%%%%%%%%%%%%%%%%%%%%%%%%%
%%%%%%%%%%%%%%%%%%%%%%%%%%%%%%%%%%%%%%%%

\begin{table*}
\begin{minipage}{175mm}
  \caption{Position centres, absolute and relative fluxes, and
    absolute and relative flux densities of the 16 sub-regions.  Flux
    densities for the Parkes 1.4 and 4.8~GHz images are presented in
    Jy. The IRIS 60~$\mu$m and 100~$\mu$m flux densities are presented
    in $10^3$ Jy. The fluxes for the \HI\ and CO are presented in $10^4$ Jy
    \kms.  The relative contribution for a sub-region (shown in
    parentheses) is given as a percentage of the total emission in the
    entire region shown in Fig.~\ref{fig:LMCmap}.}
  \label{tbl:SRpos}
  \begin{center}
    \begin{tabular}{@{}lccrrrrrrl}
   \hline
    Reg. & R.~A. & Dec. & $S_{1.4}$ & $S_{4.8}$ & $S_{60\mu m}$ &
    $S_{100\mu m}$ & $I_{\rm HI}$ & $I_{\rm CO}$ & Notes \\
         & (J2000) & (J2000) & Jy & Jy & $10^{3}$ Jy & $10^{3}$ Jy & $10^{4}$ Jy \kms & $10^{4}$ Jy \kms & \\
   \hline
    1 & 4:52:56 & $-$70:49:30  & 4.4   (1\%)  & 1.4   (0\%)  & 0.6  (1\%)  & 2.7   (1\%)  &  0.9  (2\%)  &  0.6     (1\%)   &  \\
    2 & 5:09:22 & $-$70:49:30  & 9.0   (2\%)  & 4.1   (1\%)  & 1.4  (1\%)  & 4.2   (2\%)  &  0.9  (2\%)  &  3.4     (4\%)   &  \\
    3 & 5:25:49 & $-$70:49:30  & 20.7  (4\%)  & 12.3  (4\%)  & 2.7  (3\%)  & 6.8   (3\%)  &  1.5  (4\%)  &  6.6     (7\%)   &   \\
    4 & 5:42:15 & $-$70:49:30  & 25.4  (5\%)  & 13.7  (5\%)  & 2.9  (3\%)  & 10.1  (5\%)  &  4.1  (10\%) &  12.1    (14\%)  &  \\
    5 & 4:54:29 & $-$69:28:30  & 14.0  (3\%)  & 9.2   (3\%)  & 3.8  (4\%)  & 9.1   (4\%)  &  1.5  (4\%)  &  5.0     (6\%)   &  \\
    6 & 5:09:53 & $-$69:28:30  & 21.8  (5\%)  & 13.3  (5\%)  & 6.0  (6\%)  & 12.2  (6\%)  &  1.5  (4\%)  &  8.1     (9\%)   &  \\
    7 & 5:25:18 & $-$69:28:30  & 39.4  (8\%)  & 27.0  (9\%)  & 8.1  (8\%)  & 16.0  (8\%)  &  1.6  (4\%)  &  6.3     (7\%)   & N132D \\
    8 & 5:40:42 & $-$69:28:30  & 129.6 (28\%) & 107.2 (36\%) & 36.9 (38\%) & 53.5  (26\%) &  4.9  (12\%) &  16.5    (19\%)  & 30 Dor \\
    9 & 4:55:51 & $-$68:07:30  & 8.3   (2\%)  & 4.7   (2\%)  & 2.0  (2\%)  & 5.4   (3\%)  &  1.6  (4\%)  &  3.9     (4\%)   &  \\
   10 & 5:10:21 & $-$68:07:30  & 21.0  (4\%)  & 11.5  (4\%)  & 3.5  (4\%)  & 8.5   (4\%)  &  2.1  (5\%)  &  4.7     (5\%)   &  \\
   11 & 5:24:50 & $-$68:07:30  & 34.2  (7\%)  & 24.0  (8\%)  & 7.6  (8\%)  & 15.9  (8\%)  &  2.6  (7\%)  &  8.9     (10\%)  &  \\
   12 & 5:39:20 & $-$68:07:30  & 24.7  (5\%)  & 19.0  (6\%)  & 4.8  (5\%)  & 11.2  (5\%)  &  2.3  (6\%)  &  3.5     (4\%)   & \\
   13 & 4:57:03 & $-$66:46:29  & 14.1  (3\%)  & 9.8   (3\%)  & 3.7  (4\%)  & 8.2   (4\%)  &  1.9  (5\%)  &  5.3     (6\%)   & N11 \\
   14 & 5:10:45 & $-$66:46:29  & 11.8  (3\%)  & 4.8   (2\%)  & 1.6  (2\%)  & 4.6   (2\%)  &  1.7  (4\%)  &  2.3     (3\%)   & PKS 0515-674 \\
   15 & 5:24:26 & $-$66:46:29  & 17.8  (4\%)  & 10.6  (4\%)  & 2.6  (3\%)  & 6.6   (3\%)  &  1.5  (4\%)  &  2.9     (3\%)   &  \\
   16 & 5:38:08 & $-$66:46:29  & 13.1  (3\%)  & 8.7   (3\%)  & 1.9  (2\%)  & 5.0   (2\%)  &  1.0  (3\%)  &  0.7     (1\%)   &  \\
   \hline													    
    \end{tabular}
   \end{center}
\end{minipage}
\end{table*}

%%%%%%%%%%%%%%%%%%%%%%%%%%%%%%%%
\subsection{Radio Continuum Data}
%%%%%%%%%%%%%%%%%%%%%%%%%%%%%%%%
\label{sect:RCdata}

We used a map of the 1.4~GHz radio continuum emission from the LMC
produced by combining data from the ATNF Parkes 64\,m telescope and
the 6-element Australia Telescope Compact Array (ATCA).\footnote{The
  Australia Telescope is funded by the Commonwealth of Australia for
  operation as a National Facility managed by CSIRO.}  Details of the
observations and data reduction were presented in papers by
\citet{kimetal98,kimetal03}. The angular resolution of the image is
40 arcsec, corresponding to a spatial resolution of 10~pc.  The image
suffers from ring-like artefacts at the $\sim$0.5\% level, presumably
due to errors in the primary beam model used during the mosaicking and
deconvolution process.  These errors become significant close to
bright compact sources such as the 30 Doradus region on the eastern
side of the galaxy, limiting the sensitivity that can be achieved in
these regions.\\

The 1.4~GHz image contains a large number of point sources, of which
$\ga$90\% are background AGN \citep[][Filipovic et al., in
  prep.]{marxetal97}. For the purpose of our study, it was desirable
to remove these, since they lead to a spurious decorrelation on small
scales.  We found that most of the background radio sources could be
eliminated by median filtering the image.  The filtering operation,
implemented in the GIPSY routine {\tt mfilter}, moves a
2\farcm5$\times$2\farcm5 window across a map, replacing the central
pixel value ($S_{cpix}$) with the median value of the window
($S_{med}$) if $|S_{cpix}-S_{med}| > S_{med}$ + 1 mJy.  The 1 mJy
offset prevents unnecessary filtering in noisy regions where the
median is close to zero.  The final median filtered radio continuum
map is shown in Fig.~\ref{fig:LMCmap} ({\it middle right}).  For the
whole LMC, the filter removed $\sim$10\% of the total emission in the
1.4~GHz map.  For individual sub-regions, the fraction of the total
emission removed by median filtering varies between a few percent, for
sub-regions with strong emission from the LMC itself, to $\sim$40\% in
sub-region 1. The ring artefacts surrounding extremely bright sources
such as 30 Doradus (subregion 8), N132D (subregion 7) and PKS0515-674
(subregion 14) were not removed by median filtering.\\

Lower resolution radio maps of the LMC at 1.4 and 4.8~GHz were
obtained from the Parkes surveys of \citet{haynesetal86,haynesetal91}
in order to estimate the large scale variation in the thermal fraction
of the radio emission.  The angular resolutions of the single-dish 1.4
and 4.8~GHz radio maps are 15.2 and 4.8 arcmin respectively. Further
details are presented in the paper by \citet{filipovicetal95}.

%%%%%%%%%%%%%%%%%%%%%%%%%%%%%%%%
\subsection{\HI\ Data}
%%%%%%%%%%%%%%%%%%%%%%%%%%%%%%%%
\label{sect:HIdata}

\citet{kimetal03} have combined Parkes single-dish neutral hydrogen
observations of the LMC with an ATCA aperture synthesis mosaic to
generate an \HI\ data cube for the LMC with limiting angular
resolution of 1 arcmin.  The final dataset is sensitive to spatial
scales above 15~pc across the heliocentric velocity range 190 km
s$^{-1}$ to 388 km s$^{-1}$.  The velocity resolution of the \HI\ data
is 1.65 km s$^{-1}$.  An integrated intensity map was produced from
the combined data cube by integrating over the heliocentric velocity
range of 195 to 355 \kms\ and clipping values with absolute value
$<2\sigma$. The use of clipping does bias the resulting map against
low-level emission below the clipping value, but in total the excluded
flux is only a small fraction ($\sim$8\%) of the flux in an unclipped
integrated map.

%%%%%%%%%%%%%%%%%%%%%%%%%%%%%%%%
\subsection{Molecular Gas Data}
%%%%%%%%%%%%%%%%%%%%%%%%%%%%%%%%
\label{sect:COdata}

A survey in $^{12}$CO(J=1--0) has been carried out by NANTEN, a 4m
radio telescope operated by Nagoya University at Las Campanas
Observatory in Chile.  This new survey, introduced by
\citet{fukuietal01}, has achieved a sensitivity equivalent to
$N$(H$_{2}$) $\sim 1.0 \times 10^{21}$ cm$^{-2}$, a factor of $\sim$3
better than the original survey \citep{fukuietal99,mizunoetal01}.  The
data covers an irregularly shaped area of $\sim 6^{\circ} \times
6^{\circ}$ centered on the B1950 position (5$^{\rm h}$20$^{\rm
m}$00$^{\rm s}$, $-69$\degr00\arcmin00\arcsec), and has revealed a
distribution of 259 giant molecular clouds with 27,000 observed
positions \citep{fukuietal01}.  The half-power beam width of the
telescope was 2.6 arcmin and data were taken with a 2 arcmin grid
spacing.  The spectrometers were two acousto-optical spectrometers
with 2048 channels each.  The narrow-band (NB) spectrometer has a
velocity coverage of 100 \kms\ and a resolution of 0.1 \kms, whereas
the wide-band (WB) spectrometer has a velocity coverage of 650 \kms\
and a resolution of 0.65 \kms.  Out of the 27,000 positions, 6229 were
observed with the NB spectrometer, while the rest were observed with
the newly developed WB spectrometer.\\

An integrated intensity map was produced from the NANTEN data cube as
follows.  The velocity channels were first binned to 1.95 \kms. To
isolate regions of emission, a blanking mask was defined using the
3-$\sigma$ contour of a smoothed version of the data cube; this cube
had been generated by convolution of the binned cube with a Gaussian
kernel with FWHM=$5\farcm2$. To create the integrated intensity map,
the mask was applied to the original binned cube, which was then
summed over local-standard-of-rest velocities of 200 to 300 \kms. \\

%%%%%%%%%%%%%%%%%%%%%%%%%%%%%%%%%%%
%%%%%%%%%%%%%%%%%%%%%%%%%%%%%%%%%%%
\section{Analysis}
%%%%%%%%%%%%%%%%%%%%%%%%%%%%%%%%%%%
%%%%%%%%%%%%%%%%%%%%%%%%%%%%%%%%%%%
\label{sect:analysis}

For this study, we are interested in both the structural
correspondence between images taken in different wavebands and changes
in the ratio of radio to FIR emission across the LMC. These
considerations led us to adopt two techniques for our
cross-correlation analysis: a wavelet method, which allows us to probe
morphological correlations across the whole range of spatial scales
present in each image, and a more traditional pixel-by-pixel
comparison, which is sensitive to changes in the emission ratio as a
function of position. We examine the FIR/radio ratio across the LMC in
order to compare our results with other studies of the radio-FIR
correlation in nearby disc galaxies.\\

Prior to conducting our analysis, maps of the whole LMC and the
individual sub-regions were smoothed to a common resolution. For
comparisons between the \HI\ and 1.4~GHz continuum data, the continuum
data were smoothed to 1 arcmin, i.e.\ the resolution of the
\HI\ data. For comparisons with the CO data, maps were smoothed to 3
arcmin, slightly larger than the FWHM of the NANTEN beam since the map
was not Nyquist sampled. Comparisons involving the 60~$\mu$m and
100~$\mu$m maps were less straightforward, due to the spatially
varying HIRAS resolution in response to survey coverage and
signal-to-noise.  We tested for potential resolution-dependent effects
on our results by repeating our cross-correlation analysis with a
HIRAS map smoothed to a range of resolutions between 1.5 and 5 arcmin.
This simple validation test indicated that for most sub-regions the
cross-correlation spectrum at small scales can be sensitive to changes
in resolution, but only when the variation in the resolution is quite
large (i.e.\ a factor of $\sim$2). Visual inspection of the HIRAS
data, however, suggested that the resolution of the map around 30
Doradus may be significantly lower than our nominal value of 1.3
arcmin. Further discussion of possible resolution effects on the
cross-correlation results for this sub-region is presented in
Section~\ref{sect:xcors}.\\
 
All the data preparation and analysis tasks in this study were
performed with the \textsc{Miriad}, \textsc{Gipsy} and \textsc{Yorick}
software packages
\citep{saultetal85,vanderhulstetal92,munro95}.\footnote{http://yorick.sourceforge.net}

%%%%%%%%%%%%%%%%%%%%%%%%%%%%%%%%%%%
\subsection{Pixel-by-pixel Analysis}
%%%%%%%%%%%%%%%%%%%%%%%%%%%%%%%%%%%
\label{sect:pixcmp}

The simplest measure of the correlation between two images,
$f_{1}(x,y)$ and $f_{2}(x,y)$, with the same angular resolution and
the same number of pixels is Pearson's linear correlation coefficient,
$r_{p}$. This is a direct calculation of the correlation at each pixel:
\begin{equation}
r_p = \frac{\Sigma(f_{1i}-\langle f_{1}\rangle )(f_{2i}-\langle f_{2}\rangle )}
      {\sqrt{\Sigma(f_{1i}-\langle f_{1}\rangle )^{2}\,
      \Sigma(f_{2i}-\langle f_{2}\rangle )^{2}}}
\end{equation}
Identical images should have $r_{p}=1$; images that are perfectly 
anti-correlated should have $r_{p}=-1$. The formal error on the
correlation coefficient depends on the strength of the correlation and
the number of independent pixels, $n$, in an image:
\begin{equation}
\Delta r_p = \frac{\sqrt{1-r_p^{2}}}{\sqrt{n-2}}
\end{equation}
In our case, $n=2601$ for the sub-region maps and $n=72900$ for the
whole LMC, so the formal error is small. The true error in the
correlation coefficient is dominated by systematic errors in the data,
such as calibration and zero-level uncertainties, which are not so
easily quantified.\\

The correlation coefficients for the median filtered 1.4~GHz and
60~$\mu$m emission in the whole LMC and individual sub-regions are
presented in Table~\ref{tbl:SRdata2}.  The correlation between $\log
S_{1.4}$ and $\log S_{60}$ was calculated, with flux densities expressed in
Janskys. To ensure statistical independence, one in every five pixels
along both axes were used for the calculation of $r_{p}$,
corresponding to a separation of 1.7 arcmin for the whole LMC map, and
1.6 arcmin for the sub-region maps. For the whole LMC, $r_p$=0.86,
while the median of the correlation coefficients for the sub-regions
is 0.73.  The correlation coefficients of individual sub-regions show
a large scatter about this value, ranging between 0.36 and 0.91. As
measured by Pearson's correlation coefficient, however, the
correlation between the 1.4~GHz and 60~$\mu$m (or 100~$\mu$m) emission
across the LMC is much stronger than the correlation found for other
image pairs (see Table~\ref{tbl:coefficients}).\\

\begin{figure}
\includegraphics[width=84mm]{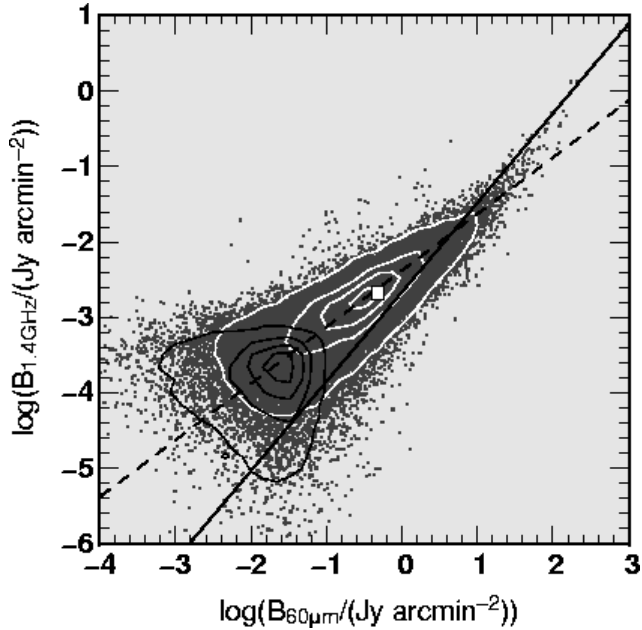}
\caption{Pixel-by-pixel correlation between the 60~$\mu$m and median
  filtered 1.4~GHz maps of the LMC. The solid line is the WLS fit to
  the data (slope = 1.18), while the dashed line is the OLS fit (slope
  = 0.75).  The white contours indicate the density of data points,
  with contour levels 5, 30, 60 and 90\% of the distribution's peak
  density. The black contours, also with contour levels at 5, 30, 60
  and 90\% of the noise's peak density, indicate the distribution that
  would be expected for Gaussian noise with RMS appropriate to the
  1.4~GHz and 60~$\mu$m images. The open square represents the average
  surface brightness for the whole LMC, calculated using the total
  ATCA+Parkes 1.4~GHz and IRIS 60~$\mu$m flux densities listed in
  Table~\ref{tbl:images}, and a total galaxy area of 7\fdg5 $\times$
  7\fdg5.}
\label{fig:pixcmpLMC}
\end{figure}

\begin{figure*}
\includegraphics[width=175mm]{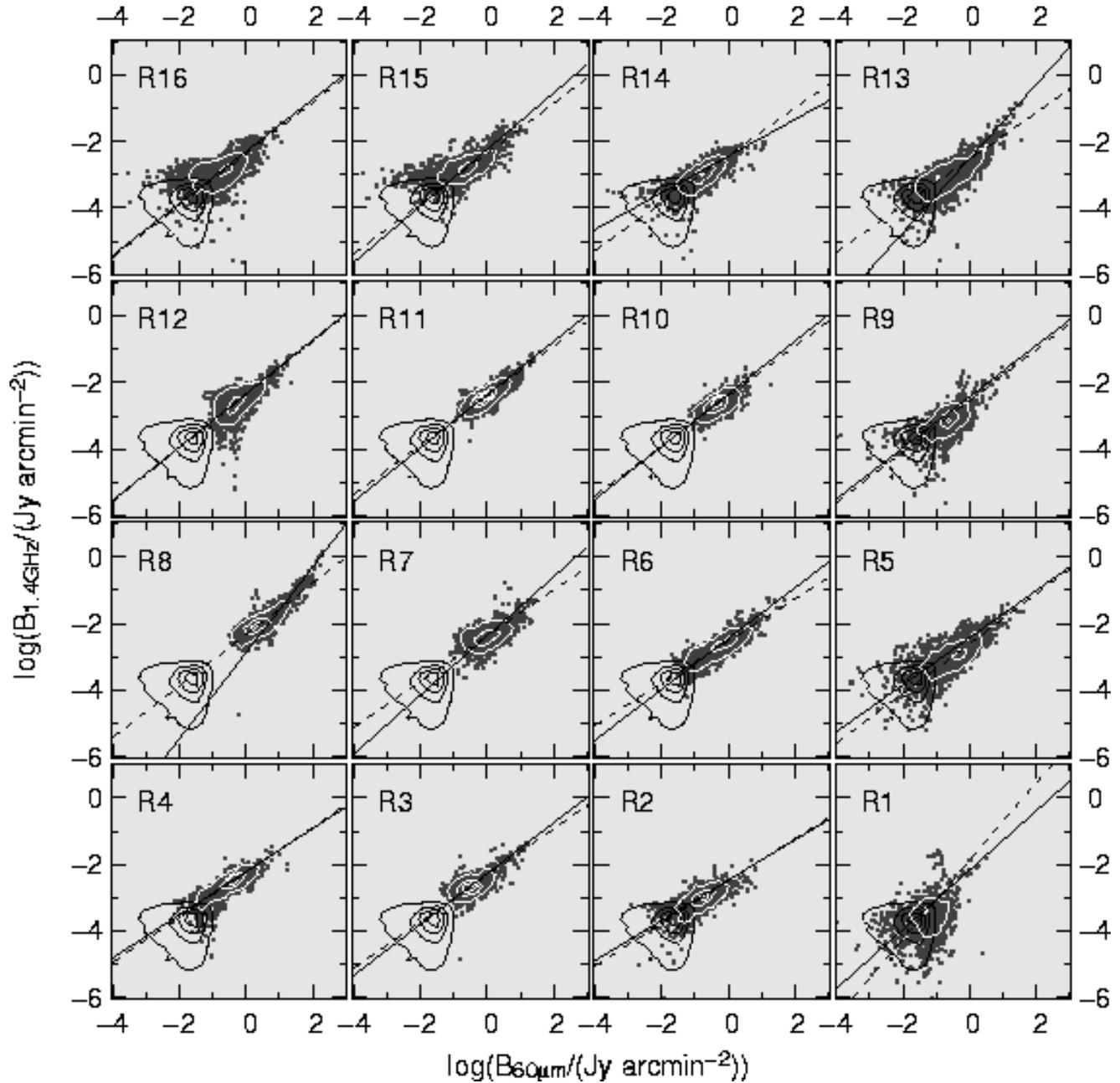}
\caption{The pixel-by-pixel correlation between the 60~$\mu$m and
  median filtered 1.4~GHz radio maps for individual sub-regions. The
  linear fits and noise contours (black) are the same as in
  Fig.~\ref{fig:pixcmpLMC}.  The white contours represent the density
  of the data points. The contour levels are 5, 30, 60 and 90\% of
  the peak density in sub-region 11.}
\label{fig:pixcmpSR}
\end{figure*}

Given that a strong correlation is present, we can consider how the
slope of the correlation varies between the individual sub-regions and
the LMC as a whole.  Scatter plots in logarithmic coordinates showing
the flux density of the 60~$\mu$m against the median filtered 1.4~GHz
radio emission for the whole LMC and the individual sub-regions are
shown in Figs.~\ref{fig:pixcmpLMC} and \ref{fig:pixcmpSR}
respectively. As for Pearson's correlation coefficient, the plots are
generated from maps that were smoothed to 1.5 arcmin resolution; the
pixels that are used to construct the scatterplots are separated along
both axes by 1.7 arcmin for the whole LMC maps, and 1.6 arcmin for the
sub-regions. As many data points can occupy the same region in a plot,
we indicate the distribution of the data points by binning the data
into a 2-D histogram and showing contours of the 2-D distribution. The
mesh used to define the contour levels consists of 60 $\times$ 60
cells, evenly distributed in logarithmic space between $10^{-7}$ and
$10^{3}$ Jy arcmin$^{-2}$ for the FIR emission, and $10^{-9}$ and
$10^{1}$ Jy arcmin$^{-2}$ for the radio emission. \\

%%%%%%%%%%%%%%%%%%%%%%%%%%%%%%%%
%%%%%%%%%%%%%%%%%%%%%%%%%%%%%%%%
%% FIGURES 2,3,4              %%
%%%%%%%%%%%%%%%%%%%%%%%%%%%%%%%%
%%%%%%%%%%%%%%%%%%%%%%%%%%%%%%%%

\begin{figure*}
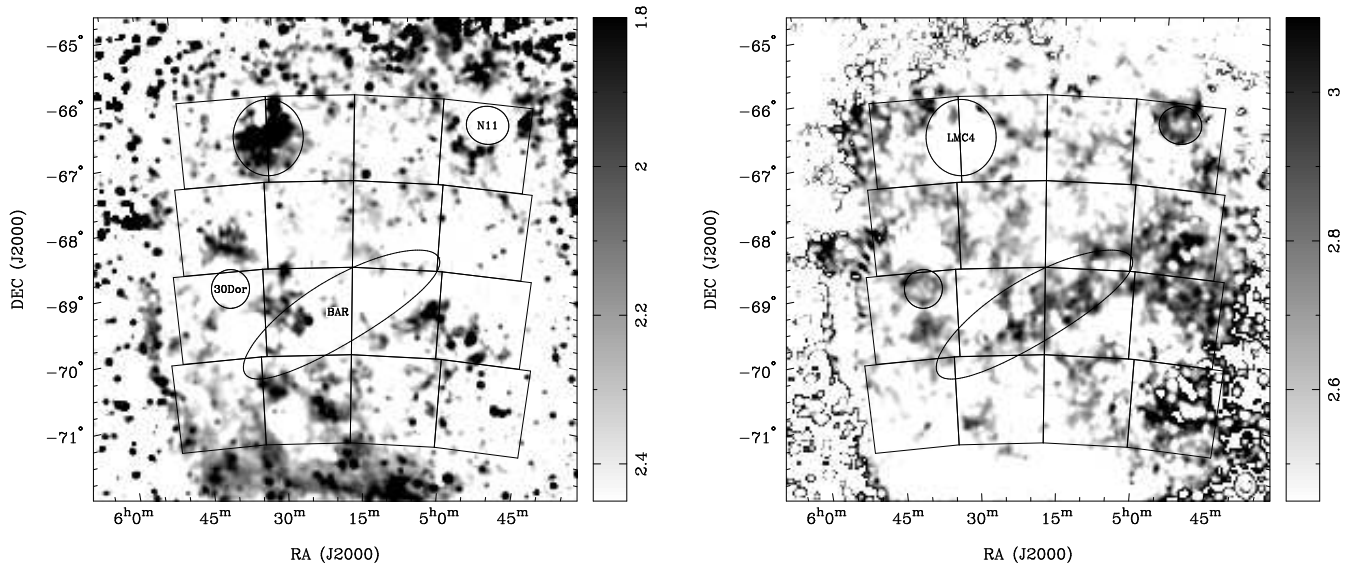

\begin{center}
\includegraphics[width=85mm]{fig5a.eps}\hfill
\includegraphics[width=85mm]{fig5b.eps}
\caption{Images of the logarithmic FIR/radio ratio, $q$, constructed
  from the combined IRIS 60 and 100~$\mu$m maps and a smoothed version
  of the ATCA+Parkes 1.4~GHz radio map.  The left panel shows regions
  where $q$ is less than its median value of 2.45, whereas the right
  panel shows regions where $q>2.45$. Major structural features of the
  LMC are indicated.}
\label{fig:qmap}
\end{center}
\end{figure*}

For both the individual sub-regions and the whole LMC, we find that
the local correlation between the radio and the 60~$\mu$m emission is
reasonably good. We used two methods for calculating a linear fit to
the pixel-by-pixel plots: weighted least squares (WLS) and the
ordinary least squares (OLS) bisector.  The WLS fit (solid line),
implemented using the Numerical Recipes {\tt fitexy} routine
\citep{press02}, assumes an uncertainty of $2.3 \times 10^{-4}$
Jy arcmin$^{-2}$ for each 1.4~GHz measurement, and $2.7 \times 10^{-2}$
Jy arcmin$^{-2}$ for each 60~$\mu$m measurement.  These uncertainities were
estimated from the RMS of a blank region of sky in the 1.4~GHz and
60~$\mu$m LMC maps at 1.5 arcmin resolution. In logarithmic space, the
high intensity pixels are thus given greater weighting due to their
smaller relative errors (note that this assumes the errors are mainly
additive rather than multiplicative). The OLS bisector (dashed line),
implemented using the {\tt slopes} routine provided by
\citet{feigelsonbabu92}, assigns equal weight to each pixel and treats
both variables symmetrically by determining the line which bisects the
standard OLS solutions of Y on X and X on Y.  Measurement
uncertainties are assumed to be unimportant compared to the intrinsic
scatter in the data.  We therefore expect that the WLS method will
provide a better fit to the high-intensity pixels, whereas the OLS
bisector will provide a better fit to the overall distribution of
pixels.  In order to prevent noise pixels contributing to our results,
a 3-$\sigma$ cut was applied to the data prior to determining the
fits.\\

For the whole LMC, the WLS method determined a slope of 1.18 to the
data, while the OLS bisector indicated a flatter slope (0.75). Similar
values are found for sub-regions 8 and 13, which contain the 30
Doradus and N11 star-forming complexes respectively. This is not
surprising, since sub-region 8 is responsible for 38\% (28\%) of the
LMC's total 60~$\mu$m (1.4~GHz) emission, and contributes
significantly to the slope of WLS fit that we determine for the galaxy
as a whole. The slopes of the WLS and OLS bisector fits in other
sub-regions (listed in Table~\ref{tbl:SRdata2} and plotted in
Fig.~\ref{fig:pixcmpSR}) show a range of values between 0.55 and
1.29. The median slope of the WLS fits is 0.78, while the median slope
of the OLS bisectors is 0.75. Contrary to a previous study of the
radio-FIR correlation in the LMC \citep{xuetal92}, our results
suggest that locally there are two independent correlations between
the radio and FIR emission, both of which are non-linear. The
correlation in the bright star-forming regions is steeper than a
linear correlation (slope$\sim$1.1--1.3), and it is
superimposed on a more general correlation that is flatter than a
linear correlation (slope$\sim$0.6--0.9). We compare our
results to the \citet{xuetal92} result in Section~\ref{sect:previous},
and discuss the variation of the WLS fit with the thermal fraction of
the radio emission in Section~\ref{sect:fthermal}. \\

Besides sub-regions 8 and 13, which contain 30~Dor and N11 and hence
have a significant population of high-intensity pixels, the WLS fit
and OLS bisector show greatest divergence in sub-region 1, where a
large fraction of the emission is close to the sensitivity limit and
the fit is therefore poorly constrained. The small congregation of
points that show a high radio to FIR ratio in sub-region 1 is due to
the bright supernova remnant 0450-70.9 \citep{mathewsonetal85}. \\

\citet{yunetal01} derived a relation between the integrated 1.4~GHz and
60~$\mu$m emission from star-forming galaxies given by 
\begin{equation}
\log (L_{1.4}/\textrm{W Hz}^{-1}) = (0.99\pm0.1)\log(L_{60}/L_{\odot})+(12.07\pm0.08)\;.
\end{equation}
We used the data in their Table 1 to re-derive this relation in flux
density units; a linear least squares fit gives
\begin{equation}
\log (S_{1.4}/\textrm{mJy}) = 0.98\log(S_{60}/\textrm{Jy})+0.86\;.
\end{equation}
These relations are shown as dashed lines in Fig.~\ref{fig:yunplot}. On
both panels, the LMC lies slightly below fits to the \citet{yunetal01}
data. Furthermore, except for sub-regions 8 and 13, the slopes that we
fit to the data points are flatter than a linear correlation.  This
causes the FIR/radio ratio to decrease for lower brightness regions,
as has been noted in four nearby spiral galaxies by
\citet{murphyetal05}. \citet{murphyetal05} attribute this
non-linearity of the local radio-FIR correlation to the diffusion of
cosmic-ray electrons (CRe$^{-}$s) away from star-forming sites,
i.e.\ an individual star-forming complex will show an infrared excess
due to local dust heating by UV photons, but it also generates an
extended non-thermal radio halo, whose radius depends on the mean age
of the CRe$^{-}$ population.\\

%%%%%%%%%%%%%%%%%%%%%%%
\subsection{FIR/radio ratio map}
%%%%%%%%%%%%%%%%%%%%%%%%
\label{sect:qmap}

In order to determine whether variations in the radio-FIR correlation
correspond to particular physical structures in the LMC, we
constructed a logarithmic FIR/radio ratio map (``$q$-map'') from the
IRIS 60 and 100~$\mu$m maps, and the ATCA+Parkes 1.4~GHz radio
map. These input maps were each smoothed to a resolution of 4.3 arcmin,
i.e.\ the natural resolution of the 100~$\mu$m image. Following the convention
of \citet{helouetal85}, $q$ is defined as
\begin{equation}
q = \log \left( \frac{\rm{FIR}}{3.75 \times 10^{12}\,\rm{W\,m^{-2}}}
\right) - \log \left( \frac{S_{1.4}}{\rm{W\,m^{-2}\,Hz^{-1}}} \right)\;.
\end{equation}
For each pixel, the total FIR surface brightness was determined from a
combination of the IRIS 60 and 100~$\mu$m maps according to Equation 6
of \citet{yunetal01}, 
\begin{equation}
{\rm FIR} = 1.26 \times 10^{-14}\,(2.58\,B_{60}+B_{100})\,\rm{W\,m^{-2}}\;,
\end{equation}
where $B_{60}$ and $B_{100}$ are the 60 and 100~$\mu$m surface
brightness in Jy pix$^{-1}$ respectively.\\

The resultant map is shown in Fig.~\ref{fig:qmap}. For clarity, the
left panel displays pixels for which $q>2.45$, and the right panel
displays pixels where $q<2.45$. Only pixels detected above the
3$\sigma$ level in the radio map were included in the calculation,
which produces an irregular-shaped mask in the final map. The median
$q$ across the whole LMC is 2.45$\pm$0.22, where 0.22 is the
semi-interquartile range. For the LMC as a whole, $q=2.55$, calculated
using the total flux density of the IRIS 60/100$\mu$m and combined ATCA+Parkes
1.4~GHz maps. For comparison, the \citet{yunetal01} galaxy sample has
a mean $q=2.34$. Within the map boundaries where we calculate $q$, it
thus appears that either the radio emission in the LMC is slightly
underluminous compared to the value that we expect from the
\citet{yunetal01} correlation, or that the FIR emission is slightly
overluminous (see Fig.~\ref{fig:pixcmpLMC}). \\

In contrast to some high-resolution $q$-maps of nearby disc galaxies
\citep[see e.g.][]{murphyetal05}, the most striking structures in the
$q$-map of the LMC do not correspond to the brightest structures in
the input FIR and radio maps (see Fig.~\ref{fig:qmap}). However,
inspection of the $q$-map shows that the variation in the FIR/radio
ratio across the LMC is spatially organized to some extent. In
particular, the stellar bar region and the western edge of the LMC
show elevated FIR/radio ratios, while the eastern side of the LMC
shows numerous patches of low $q$ values. The supershell LMC4 in the
north-east is the largest of these low $q$ regions
\citep{meaburn80}. Both 30 Doradus and N11 have a pronounced ring-like
morphology in the $q$-map.  The local peak in the original radio and
FIR maps is surrounded by a ring of enhanced $q$ values. An earlier
low-resolution study by \citet{kleinetal89} has reported that LMC
\HII\ regions have an average $q=2.78$. Our data suggests that at
higher angular resolution, the $q$ value of an HII region can be
separated into two distinct components: a compact central region where
the FIR and (mostly thermal) radio emission are high but $q$ is quite
close to the mean $q$ across the whole LMC, and a ring of enhanced $q$
values that arises through the different scale-lengths of the FIR and
radio emission associated with the \HII\ region. This morphology is
consistent with the statistical detection of a ``dip-and-ring''
structure around LMC \HII\ regions by \citet{xuetal92}.

%%%%%%%%%%%%%%%%%%%%%%%%%%%%%%%%%%%
\subsection{Wavelet Analysis}
%%%%%%%%%%%%%%%%%%%%%%%%%%%%%%%%%%%
\label{sect:wavelets}

Wavelet analysis involves the convolution of an image with a family of
self-similar basis functions that depend on scale and location. It can
be useful to consider the wavelet transform as the general case of the
Fourier transform, where the oscillatory functions are localised in
both time and frequency.  The family of basis functions is generated by
dilating and translating a mother function, which is called the
analysing wavelet.  In this paper we make use of the two-dimensional
continuous wavelet transform, which can be written as
\begin{equation}
W(a,\mathbf{x})=w(a) \int_{-\infty}^{\infty} \int_{-\infty}^{\infty} 
f(\mathbf{x^\prime}) \psi^{\ast}\left(
\frac{\mathbf{x^\prime}-\mathbf{x}}{a} 
\right) \,\mathrm{d}\mathbf{x^\prime}
\label{eqn:wavdef}
\end{equation}
Here $f(\mathbf{x})$ is a two-dimensional function, such as an image,
$\psi^{\ast}$ is the complex conjugate of the dilated and translated
analysing wavelet, and $w(a)$ is a normalising function. $a$ is the
scale size of the wavelet. Note that the term ``continuous wavelet
transform'' is used in the literature to distinguish this transform
from a ``discrete'' transform which employs an orthogonal set of basis
functions; however, there is no fundamental difficulty applying the
former to discretely sampled data such as astronomical images.\\

Our use of wavelet transforms follows closely the work of
\citet{fricketal01}, who pioneered their application to galaxy images.
Our analysing wavelet is their ``Pet Hat'' which can be written in
Fourier space as
\begin{equation}
\hat{\psi}(\mathbf{k}a) = \left \{ \begin{array} 
{r@{\quad:\quad}l} 
C(a) \cos^2 \left ( \frac{\pi}{2}\log_2 2|\mathbf{k}|a \right ) & 
\frac{1}{4a} < |\mathbf{k}| < \frac{1}{a} \\ 
0 &  |\mathbf{k}| < \frac{1}{4a}, |\mathbf{k}| > \frac{1}{a}
\end{array} \right.
\end{equation}
This function picks out an annulus in the Fourier plane centred at
$k$=$(2a)^{-1}$, and provides both simplicity and good separation of
scales.  The normalising factor $C(a)$ is chosen to provide unit total
flux density in the wavelet image (or unit total flux, in the case of
the \HI and CO data). $C(a)$ is evaluated numerically to be
$1.065a/\delta$, where $a$ and $\delta$ are the scale size (defined
below) and pixel size expressed in radians.\\

\begin{figure}
\includegraphics[width=85mm]{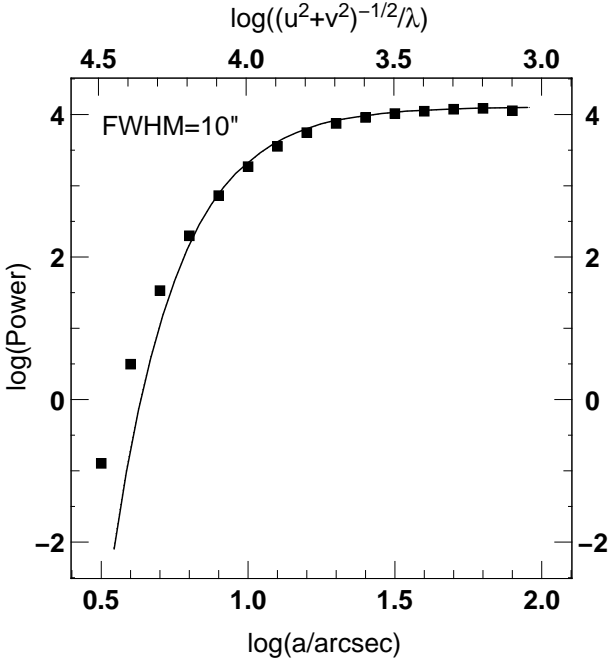}
\caption{The wavelet (filled squares) and Fourier (solid line) power spectra for a
  10\arcsec\ (FWHM) Gaussian.}
\label{fig:gauss10spec}
\end{figure}

Throughout this paper we adopt the convention that an angular scale
size $a$ in radians corresponds to a spatial frequency of $(2a)^{-1}$
wavelengths.  This is based on the observation that a positive,
Gaussian-like structure will tend to couple with half of a sine wave
rather than a full period, however it differs from the usual
convention by a factor of 2.  As an illustrative example, we show in
Fig. \ref{fig:gauss10spec} the wavelet and Fourier spectra for a
10\arcsec\ (FWHM) Gaussian; the spatial frequency is shown on the top
axis and the corresponding spatial scales are shown on the bottom
axis.  In Table~\ref{tbl:SRdata2} and the following sections, we
express $a$ in arcseconds, which are a more appropriate unit for the
size scales of interest here. The Fourier spectrum is defined in the
usual way as $E_f(k) = |\hat{f}(\mathbf{k})|^2$.\\

%%%%%%%%%%%%%%%%%%%%%%%%%%%%%%%%
%%%%%%%%%%%%%%%%%%%%%%%%%%%%%%%%
%% FIGURE 5                   %%
%%%%%%%%%%%%%%%%%%%%%%%%%%%%%%%%
%%%%%%%%%%%%%%%%%%%%%%%%%%%%%%%%

Expressing the convolution of Equation~\ref{eqn:wavdef} in terms of
Fourier transforms, the {\it wavelet filtered image} is calculated as
\begin{equation}
W(a,\mathbf{x}) = \frac{1}{N^2} \sum_{(k_1,k_2)} \hat{f}(\mathbf{k})\,
\psi^{\ast}(\mathbf{k}a)\, e^{-2\pi i(\mathbf{k}\cdot\mathbf{x})/N}\;,
\end{equation}
where $N$ is the linear dimension of the image in pixels, and the
{\it wavelet spectrum} as
\begin{equation}
E(a) = \sum_{(x_1,x_2)} |W(a,\mathbf{x})|^2\;.
\end{equation}
This spectrum, which can be thought of as a smoothed version of the
power spectrum, reveals the dominant energy scales in an image.\\

Finally we define the {\it wavelet cross-correlation coefficient} as
\begin{equation}
r_w(a) = \frac{\sum \sum W_1(a,\mathbf{x})W_2(a,\mathbf{x})}
{[E_1(a)E_2(a)]^{1/2}}
\end{equation}
where the subscripts refer to two images of the same size.  This is
the wavelet analogue of the Fourier cross power spectrum.  Following
\citet{fricketal01}, we assign an uncertainty of
\begin{equation}
\Delta r_w(a) = \frac{\sqrt{1-r_w^2}}{\sqrt{n-2}}
\end{equation}
to this coefficient, where $n = (L/a)^2$ and $L$ is the linear size of
the image.

%%%%%%%%%%%%%%%%%%%%%%%%%%%%%%%%
%%%%%%%%%%%%%%%%%%%%%%%%%%%%%%%%
%% TABLE 3      %%
%%%%%%%%%%%%%%%%%%%%%%%%%%%%%%%%
%%%%%%%%%%%%%%%%%%%%%%%%%%%%%%%%

\begin{table}
  \caption{Summary of the wavelet and pixel-by-pixel correlations
    between the median filtered 1.4~GHz and the 60~$\mu$m images for
    the whole LMC and individual sub-regions.  Columns 2 and 3 list
    the angular (arcsec) and linear (parsec) spatial scales at which
    the wavelet cross-correlation falls below 0.75.  Column 3 lists
    the Pearson's correlation coefficient, $r_{p}$. Columns 4 and 5
    list the WLS and OLS-bisector slopes. Column 6 lists the thermal
    fraction $f_{th}$ of the radio emission at 1.4~GHz, assuming a
    non-thermal spectral index of $-0.7$.}
  \begin{center}
    \begin{tabular}{@{}lcccccc}
    \hline
    Region & log($a$/arcsec)  & Scale (pc)  &  $r_{p}$ & slope$_{WLS}$ & slope$_{OLS}$ & $f_{th}$\\% & $f_{th}, \alpha=-0.9$ & $\alpha_{60}$& $\alpha_{100}$& $\alpha_{RC}$& $\alpha_{mfRC}$& $\alpha_{HI}$& $\alpha_{CO}$ \\
    \hline
  LMC & 2.3 & 50     & 0.86  & 1.18 & 0.75 & 0.45 \\%& 0.54 &   2.08 &   1.89 &  1.35  &  1.53 &  2.65 &  1.27  \\
    1 & 3.0 & $>$240 & 0.36  & 0.89 & 1.17 & $<$0 \\%& $<$0 &   2.12 &   1.38 &  0.77  &  2.11 &  2.43 &  1.74  \\
    2 & 2.7 & 120    & 0.57  & 0.60 & 0.64 & 0.07 \\%& 0.23 &   1.02 &   0.84 &  0.88  &  1.83 &  2.31 &  0.58  \\
    3 & 2.4 & 60     & 0.79  & 0.76 & 0.70 & 0.37 \\%& 0.47 &   2.30 &   2.22 &	 1.06  &  2.23 &  2.47 &  1.53  \\
    4 & 3.0 & $>$240 & 0.62  & 0.64 & 0.68 & 0.26 \\%& 0.38 &   1.34 &   1.64 &  0.90  &  1.65 &  2.68 &  1.66  \\
    5 & 2.3 & 50     & 0.72  & 0.71 & 0.75 & 0.50 \\%& 0.58 &   1.25 &   0.93 &  0.76  &  1.71 &  2.69 &  0.94  \\
    6 & 2.2 & 40     & 0.80  & 0.76 & 0.63 & 0.41 \\%& 0.51 &   1.50 &   1.07 &  1.09  &  1.55 &  2.59 &  0.95  \\
    7 & 3.0 & $>$240 & 0.48  & 0.89 & 0.69 & 0.57 \\%& 0.64 &   2.04 &   1.73 &  0.38  &  0.94 &  2.36 &  0.80  \\
    8 & 2.2 & 40     & 0.89  & 1.29 & 0.77 & 0.87 \\%& 0.90 &   2.82 &   2.44 &	 1.79  &  1.88 &  2.79 &  1.61  \\
    9 & 2.2 & 40     & 0.77  & 0.77 & 0.77 & 0.33 \\%& 0.44 &   1.23 &   1.26 &  0.87  &  1.46 &  2.60 &  1.47  \\
   10 & 2.8 & 150    & 0.58  & 0.79 & 0.75 & 0.27 \\%& 0.40 &   2.31 &   2.27 &  0.97  &  2.19 &  2.79 &  1.48  \\
   11 & 2.1 & 30     & 0.91  & 0.80 & 0.74 & 0.60 \\%& 0.67 &   2.22 &   3.20 &	 1.71  &  2.12 &  2.76 &  2.22  \\
   12 & 1.9 & 20     & 0.86  & 0.80 & 0.80 & 0.75 \\%& 0.79 &   1.57 &   1.34 &	 1.08  &  1.61 &  2.68 &  0.79  \\
   13 & 2.1 & 30     & 0.85  & 1.11 & 0.70 & 0.58 \\%& 0.65 &   1.62 &   2.81 &	 1.22  &  1.29 &  2.66 &  1.67  \\
   14 & 2.9 & 150    & 0.55  & 0.55 & 0.71 & $<$0 \\%& 0.14 &   1.22 &   1.64 &  0.07  &  1.63 &  2.49 &  1.05  \\
   15 & 2.6 & 100    & 0.72  & 0.85 & 0.75 & 0.38 \\%& 0.48 &   2.13 &   2.55 &	 2.32  &  3.03 &  3.14 &  1.22  \\
   16 & 2.3 & 50     & 0.73  & 0.78 & 0.77 & 0.53 \\%& 0.61 &   1.76 &   1.31 &	 1.31  &  1.91 &  2.54 &  0.95  \\
  \hline
  \end{tabular}
\end{center}
\label{tbl:SRdata2}
\end{table}

%%%%%%%%%%%%%%%%%%%%%%%%%%%%%%%%
%%%%%%%%%%%%%%%%%%%%%%%%%%%%%%%%
%% FIGURE 6                   %%
%%%%%%%%%%%%%%%%%%%%%%%%%%%%%%%%
%%%%%%%%%%%%%%%%%%%%%%%%%%%%%%%%

\begin{figure*}
\includegraphics[width=175mm]{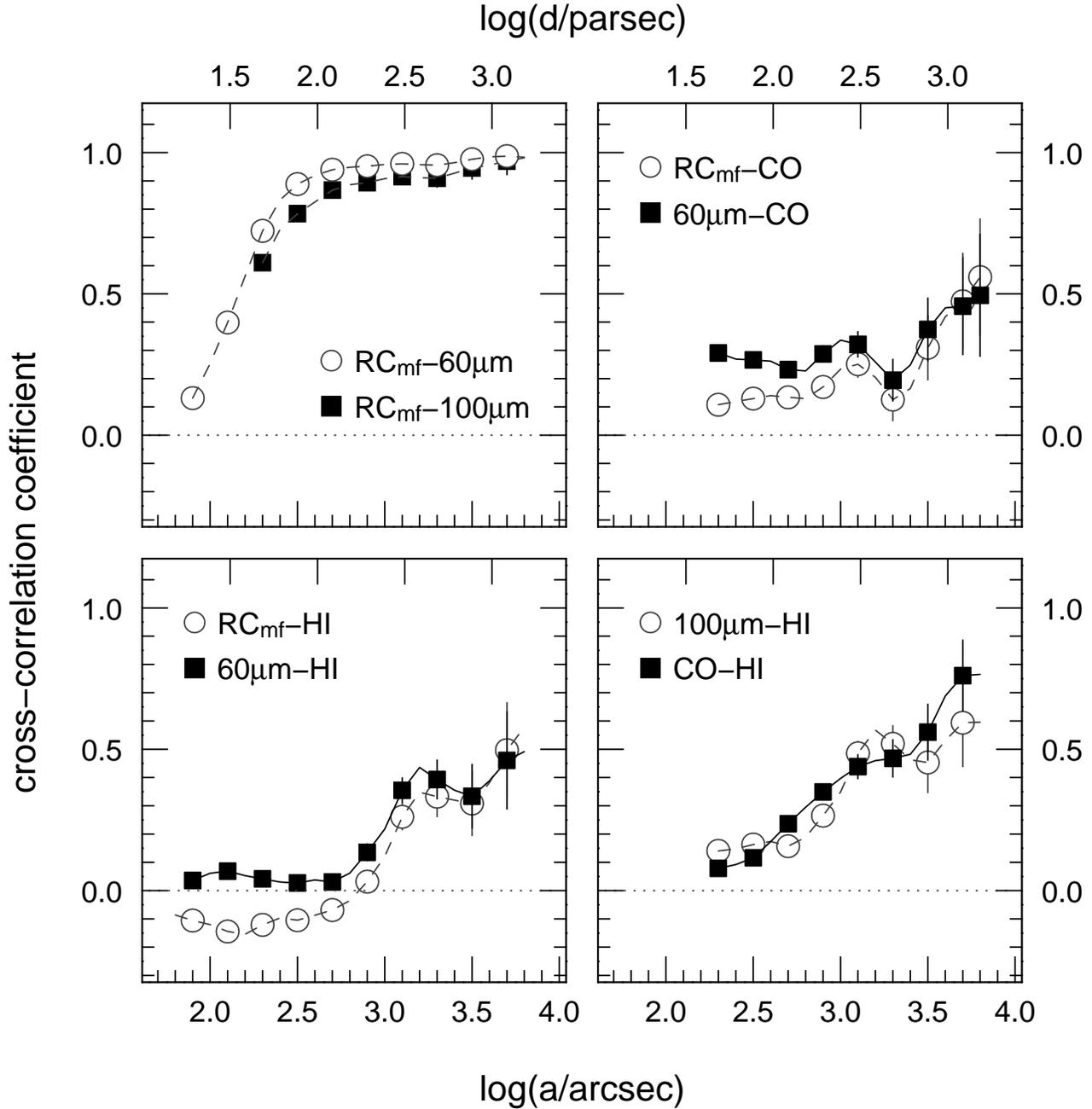}
\caption{Wavelet cross-correlation spectra for emission from the whole
  LMC. {\it Top Left.} Median filtered 1.4~GHz radio vs.\ 60~$\mu$m
  emission and median filtered 1.4~GHz radio vs.\ 100~$\mu$m emission.
  {\it Top Right.}  Median filtered 1.4~GHz radio vs.\ integrated CO
  emission and 60 $\mu$m vs.\ CO emission.  {\it Bottom Left.} Median
  filtered 1.4~GHz radio vs.\ integrated \HI\ emission and 60 $\mu$m
  vs.\ \HI\ emission.  {\it Bottom Right.} 100~$\mu$m
  vs.\ \HI\ emission and CO vs.\ \HI\ emission. For all panels, the
  wavelet scale is indicated on the bottom axis, and the corresponding
  spatial scale (for an assumed LMC distance of 50.1~kpc) is shown on
  the top axis.}
\label{fig:xcorLMC}
\end{figure*}

%%%%%%%%%%%%%%%%%%%%%%%%%%%%%%%%%%%
%%%%%%%%%%%%%%%%%%%%%%%%%%%%%%%%%%%
\subsection{Wavelet Cross-correlations}
%%%%%%%%%%%%%%%%%%%%%%%%%%%%%%%%%%%
%%%%%%%%%%%%%%%%%%%%%%%%%%%%%%%%%%%
\label{sect:xcors}

The wavelet cross-correlation spectrum was calculated for each of the
following map pairs: (a) radio and 60~$\mu$m emission; (b) radio and
100~$\mu$m emission; (c) radio and CO emission; (d) 60~$\mu$m and CO
emission; (e) radio and \HI\ emission; (f) 60~$\mu$m and
\HI\ emission; (g) 100~$\mu$m and \HI\ emission; (h) CO and
\HI\ emission.  The median filtered 1.4~GHz map was used for the
comparisons involving radio emission. We calculate the
cross-correlation spectra on wavelet scales $\log(a) = 1.8-3.8$ for
pair (e), $\log(a) = 1.9-3.8$ for pairs (a) and (f), and $\log(a) =
2.3-3.8$ for pairs (b), (c), (d), (g) and (h). For an LMC distance of
50.1~kpc, the range of spatial scales that we probe with all eight
cross-correlation spectra is $0.05-1.5$~kpc, extending down to
0.02~kpc for the highest resolution datasets. The smallest scale that
we can study is limited by the resolution of the data. The largest
scale depends on the size of the field, and is determined such that
the width of each map is equal to or greater than four times the
largest wavelet scale. The results for all eight pairs are shown in
Fig.~\ref{fig:xcorLMC} for the whole LMC.  Figs.~\ref{fig:xcorSR1} and
\ref{fig:xcorSR2} show the results for pairs (a), (c), and (e) for the
individual sub-regions, as well as the wavelet cross-correlation
between the original (unfiltered) radio image with the 60~$\mu$m
emission. For the cross-correlation spectra of the sub-regions, the
largest wavelet scale that we study is $\log(a) = 3.0$, corresponding
to a spatial scale of $\sim$0.25~kpc.  \\

\subsubsection{Cross-correlations between radio and FIR emission}

For the whole LMC, it is clear that the correlation between the
1.4~GHz radio emission and the 60 or 100~$\mu$m dust emission
(Fig.~\ref{fig:xcorLMC}, {\it top left}) is very strong, and
significantly better at all spatial scales than the correlations that
involve the molecular or atomic gas. \citet{fricketal01} regard values
higher than 0.75 to be indicative of an excellent correlation between
two images: by this measure, the correlation between the radio and
60~$\mu$m emission in the LMC remains strong down to spatial scales of
$\sim$50~pc (210 arcsec). The correlation between the 1.4~GHz radio
and the 100~$\mu$m dust emission is marginally worse across all
scales, falling below 0.75 at $\sim$60~pc (250 arcsec).  Nonetheless,
it still appears that, averaged over the entire LMC, the radio-FIR
correlation is maintained on scales less than the scale height of the
gaseous disc, which has been estimated to be between 80--100 pc
\citep{elmegreenetal01} and 180 pc \citep{kimetal99}.\\

\begin{figure*}
\includegraphics[width=175mm]{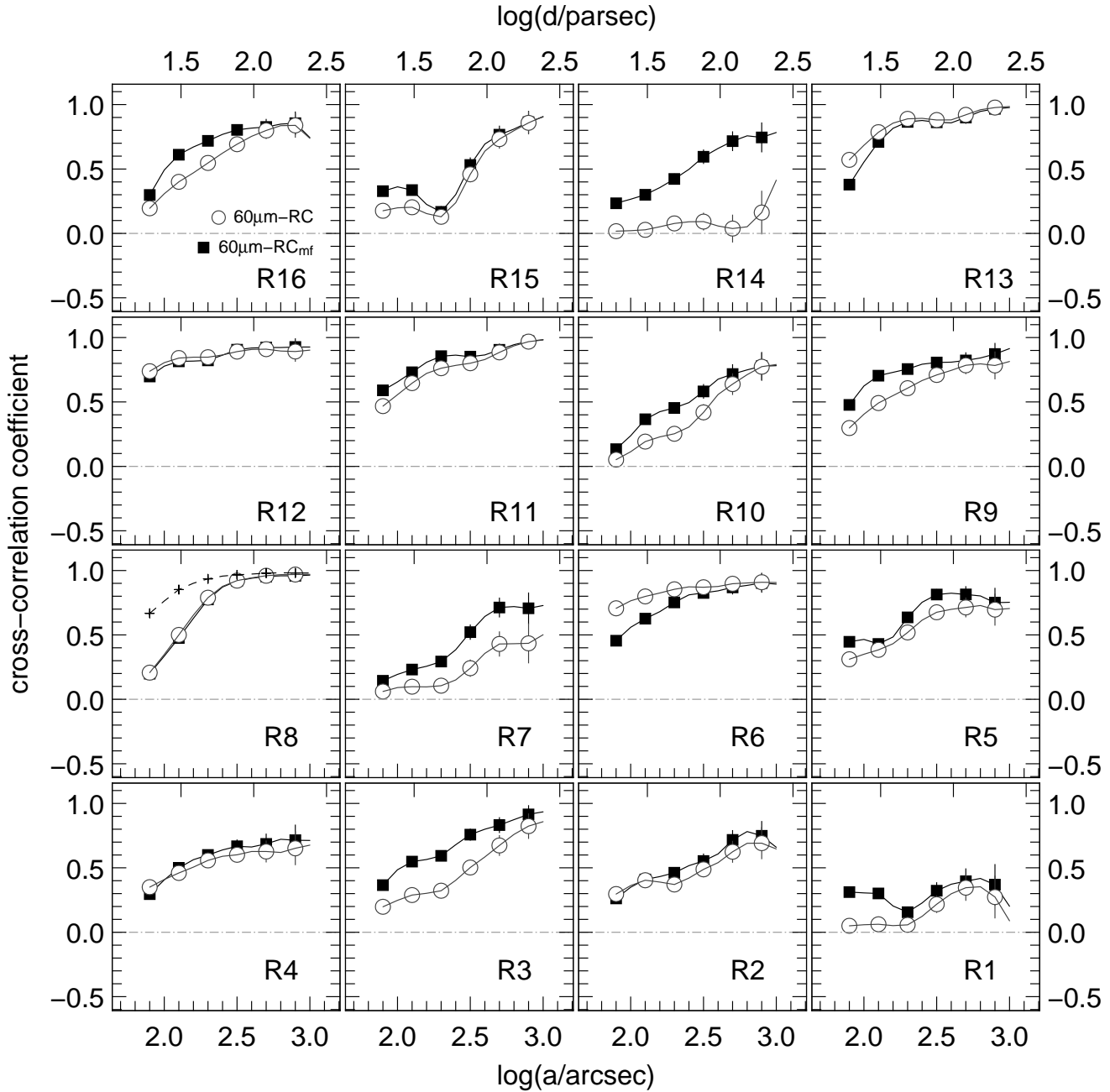}
\caption{Wavelet cross-correlation spectra in individual sub-regions
  for the 1.4~GHz radio vs.\ HIRAS 60~$\mu$m emission and median
  filtered 1.4~GHz radio vs.\ HIRAS 60~$\mu$m emission.  For
  sub-region 8, we also plot the wavelet cross-correlation spectrum
  for the 1.4~GHz radio vs.\ HIRES 60~$\mu$m emission (crosses). The
  discrepancy between the cross-correlation results for the HIRAS and
  HIRES data suggests that resolution effects may be responsible for
  the poor correlation between the HIRAS 60~$\mu$m and the 1.4~GHz
  radio emission on small scales.}
\label{fig:xcorSR1}
\end{figure*}

Since the emission of the LMC as a whole is strongly influenced by the
30 Dor region, we have also compared the median filtered 1.4~GHz and
the 60 $\mu$m emission for individual sub-regions across the LMC
(Fig.~\ref{fig:xcorSR1}, {\it filled squares}).  The spatial scale at
which the correlation falls below 0.75 for each sub-region is listed
in Table~\ref{tbl:SRdata2}.  Generally the correlation remains
excellent, with half of the sub-regions showing a strong correlation
down to spatial scales of $<$50~pc ($\log(a) \sim 2.3$).  However, in
several of the other sub-regions the correlation breaks down on
much larger scales ($>$100 pc, corresponding to $\log(a) \sim 2.6$).
We demonstrate in Section~\ref{sect:fthermal} that there is an inverse
relationship between the size scale of the correlation breakdown and
the thermal fraction of the radio emission.  Thus, the excellent
correlation between the radio and 60~$\mu$m emission that we see in
sub-regions such as 8 and 12 may be dominated by a correlation between
the warm dust and the thermal radio emission, which does not drive the
correlation between the total radio and FIR flux density of spiral galaxies.\\

Our results for the wavelet cross-correlations are fairly robust
whether or not median filtering is used to suppress compact point
sources.  When the original (unfiltered) radio map is cross-correlated
with the 60~$\mu$m map, the correlation generally degrades
(Fig.~\ref{fig:xcorSR1}, {\it open circles}), especially for fields at
the edge of the LMC where the total emission is low and fewer of the
compact point sources are presumably intrinsic (e.g. sub-regions 3, 9,
and 14).  However, the qualitative trends remain the same, since the
compact sources usually dominate the emission only on the smallest
wavelet scales.  In a few cases (e.g. sub-regions 6 and 13) the median
filtering appears to worsen the correlation, by removing emission from
compact sources that have bright 60~$\mu$m counterparts and are thus
likely to be \HII\ regions.\\

On the other hand, an incorrect assumption about the
  resolution of the input images (e.g., the HIRAS-processed IRAS
  images) could cause the correlation to break down on larger scales
  than it should.  This is especially a concern in sub-region 8, which
  covers 30 Doradus.  Wavelet power spectra for this sub-region show a
  steep decline below a scale of $\sim$250 arcsec, whereas the power
  spectra for other sub-regions tend to steepen at smaller
  characteristic scales ($\sim$120 arcsec), slightly larger than our
  estimate of the HIRAS resolution.  Using an IRAS image processed
  with the HIRES deconvolution method (kindly provided to us by
  J. Surace at IPAC) results in a much stronger wavelet
  cross-correlation that exceeds 0.75 down to scales of 100 arcsec
  (Fig.~\ref{fig:xcorSR1}, dashed line in R8). HIRES is an alternative
  re-processing of the IRAS survey, which, like HIRAS, aims to improve
  on the nominal resolution of the original IRAS survey data. HIRES
  uses the Maximum Correlation Method rather than the Pyramid Maximum
  Entropy algorithm used by HIRES, and achieves a resolution of
  $\sim$1 arcmin for the 60$\mu$m band. The discrepancy between the
  results of the wavelet cross-correlation for the HIRAS and HIRES
  data in sub-regio 8 means that we cannot rule out the possibility
  that the radio-FIR correlation remains strong on even smaller scales
  than we deduce here, especially in the 30 Dor region, which
  constitutes a large fraction of the galaxy's emission.  Future
  high-resolution data sets from {\it Spitzer} should be able to
  overcame these difficulties with the IRAS data.\\

\subsubsection{Cross-correlations between other wavebands}

Cross-correlations involving the CO and \HI\ images for the whole LMC
are shown in the last 3 panels of Fig.~\ref{fig:xcorLMC}.  For each
image pair, the value of the wavelet cross-correlation at wavelet
scales $\log (a) = 2.0$, 2.5, 3.0 and 3.5 (corresponding to spatial
scales of 25, 80, 245 and 770 pc) is listed in
Table~\ref{tbl:coefficients}.  The relative weakness of these
correlations compared to the radio-FIR correlations is immediately
apparent.  Indeed, the strength of the radio-FIR correlation is such
that correlating one of the gas tracers with either the radio or FIR
emission yields basically the same result.  There is also a strong
resemblance between the 100~$\mu$m-\HI\ and CO-\HI\ correlations which
is not so easily explained, given the lack of a strong correlation
between the 100~$\mu$m and CO emission.  One possibility is that both
the 100~$\mu$m and CO emissions reflect the typical sizes of
star-forming regions, but in different evolutionary states, whereas
the \HI\ emission is dominated by much larger structures.\\

\begin{figure*}
\includegraphics[width=175mm]{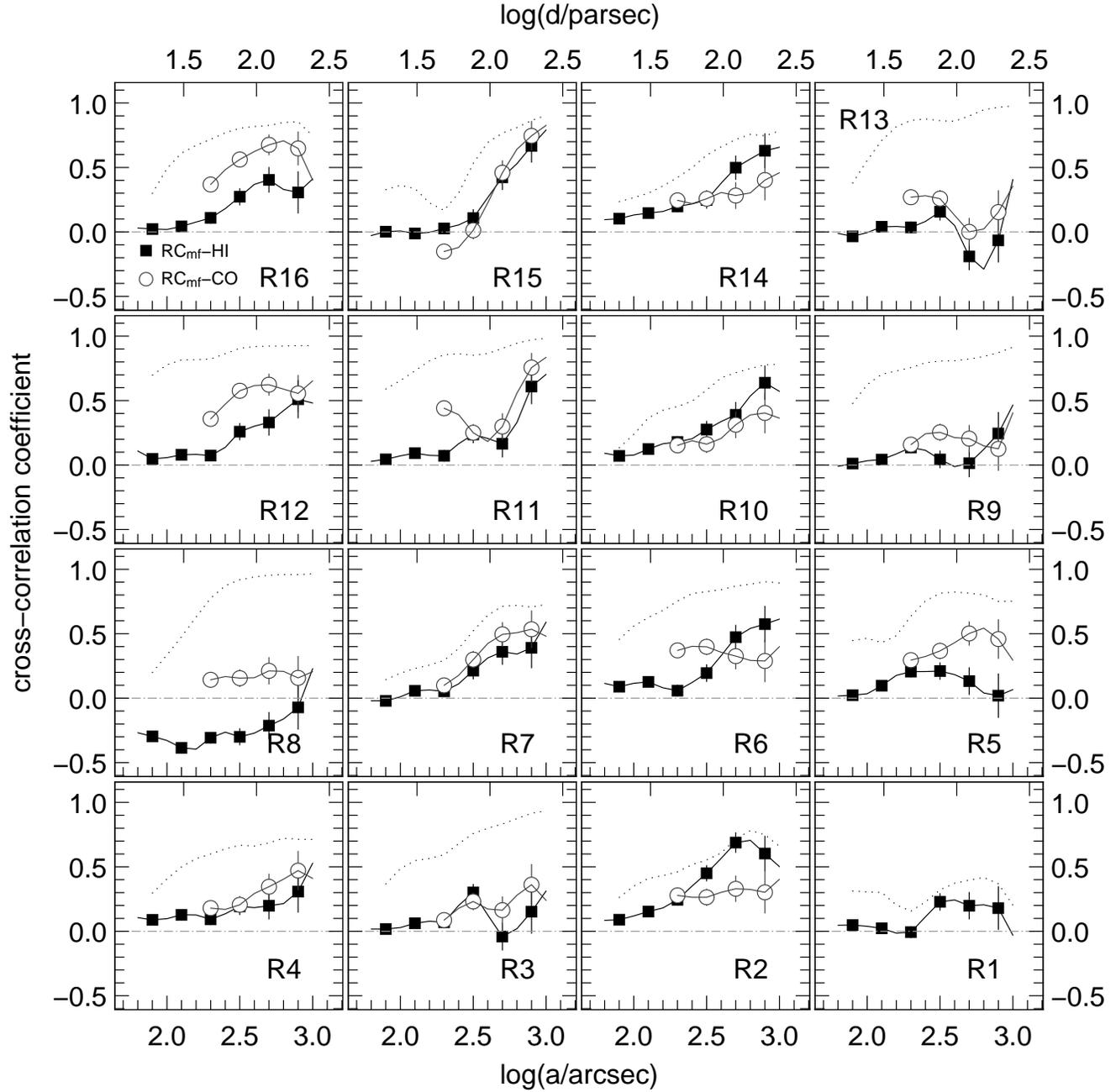}
\caption{Wavelet cross-correlation spectra in individual sub-regions
  for the integrated \HI\ vs.\ 1.4~GHz emission and
  integrated CO vs.\ 1.4~GHz emission.  The dotted
  lines show the cross-correlation spectra for the median filtered
  1.4~GHz radio vs.\ 60~$\mu$m emission from Fig.~\ref{fig:xcorSR1}.
  Note that sub-region 1 is poorly covered by our CO map.}
\label{fig:xcorSR2}
\end{figure*}

Within individual sub-regions, cross-correlations of the median
filtered 1.4~GHz with the CO and \HI\ are shown in
Fig.~\ref{fig:xcorSR2}.  As expected, for most regions these
correlations are much weaker than the cross-correlation of 60~$\mu$m
with 1.4 GHz radio.  Near 30 Doradus
(sub-region 8), the integrated \HI\ emission is weakly anti-correlated
($-0.4 < r_{w} < 0$) with the FIR and radio emission across large and
small scales.  There is also a weak anti-correlation between the
\HI\ and radio emission at $\log (a) \sim 2.7-2.8$ for the field
containing N11 (sub-region 13); this localised anti-correlation
corresponds to the dense, clumpy ring of molecular and atomic gas that
surrounds the bright 60~$\mu$m and radio source at the centre of the
N11 star-forming region.\\

%%%%%%%%%%%%%%%%%%%%%%%%%%%%%%%%
%%%%%%%%%%%%%%%%%%%%%%%%%%%%%%%%
%% FIGURES 7 and 8            %%
%%%%%%%%%%%%%%%%%%%%%%%%%%%%%%%%
%%%%%%%%%%%%%%%%%%%%%%%%%%%%%%%%

%%%%%%%%%%%%%%%%%%%%%%%%%%%%%%%%%%%
%%%%%%%%%%%%%%%%%%%%%%%%%%%%%%%%%%%
\subsection{Decomposition of the Radio Continuum Emission}
%%%%%%%%%%%%%%%%%%%%%%%%%%%%%%%%%%%
%%%%%%%%%%%%%%%%%%%%%%%%%%%%%%%%%%%
\label{sect:fthermal}

Previous studies of spiral galaxies have shown that the local
FIR-radio correlation is constituted by a correlation between the warm
dust and the thermal radio emission, and between the cool dust and the
non-thermal radio emission \citep{hoernesetal98}. In order to
investigate whether the good correlation on small spatial scales that
we found for some sub-regions is dominated by a correlation between
the warm dust and thermal radio emission, we used Parkes maps of the
LMC at 1.4 and 4.8~GHz to estimate the thermal fraction of the radio
emission across the LMC.\\

\begin{figure}
\includegraphics[width=80mm]{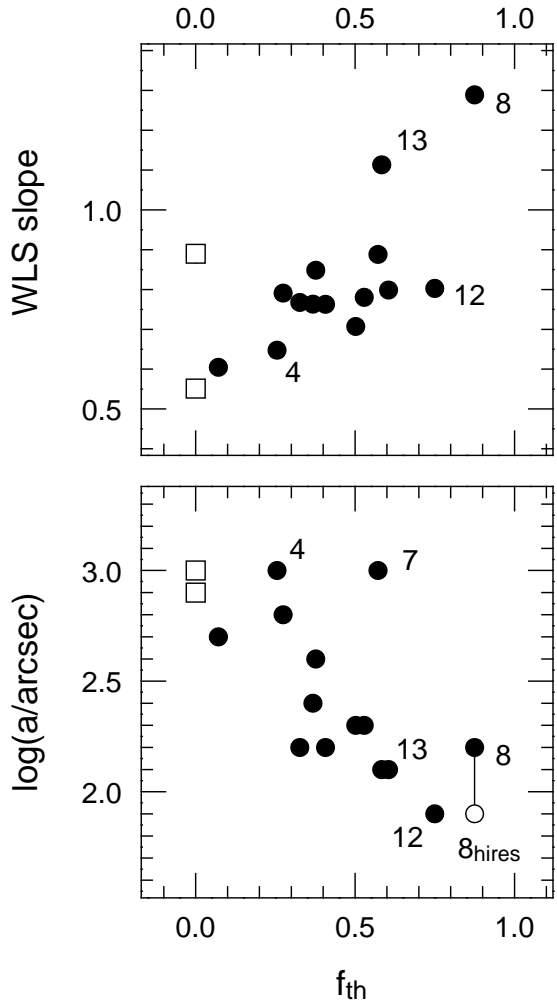}
\caption{{\it Top.} The WLS slope of the radio vs.\ 60\,$\mu$m
  relation, plotted against the thermal fraction at
  1.4~GHz. Sub-regions where our decomposition method produces thermal
  fractions less than 0 are plotted as $f_{th}=0$ with open squares.
  {\it Bottom.} The wavelet scale at which the radio-FIR wavelet
  cross-correlation falls below 0.75, plotted against the thermal
  fraction of the radio emission at 1.4~GHz. The open circle indicates
  the result for sub-region 8 using the HIRES 60$\mu$m data. }
\label{fig:SRbreakdown}
\end{figure}

First, the single-dish 1.4 and 4.8 GHz radio maps were smoothed to a
common resolution of 15\farcm2. We then estimated the thermal fraction
of the radio emission in each sub-region by applying equation 1 from
the paper by \citet{niklasetal97} to the total flux density of each
sub-region at 1.4 and 4.8 GHz,
\begin{equation}
\frac{S_{4.8}}{S_{1.4}}=f_{th} \left (
\frac{\nu_{4.8}}{\nu_{1.4}}\right )^{-\alpha_{th}} + 
f_{nth} \left ( \frac{\nu_{4.8}}{\nu_{1.4}}\right )^{-\alpha_{nth}}\;.
\end{equation}
Here, $f_{th}$ is the thermal fraction of the radio emission at
1.4~GHz, $f_{nth} = 1-f_{th}$ is the non-thermal fraction, and
$S_{1.4}$ and $S_{4.8}$ are the flux desnsities of each sub-region at
1.4 and 4.8 GHz respectively.  For the thermal spectral index, we
adopted $\alpha_{th}$=0.1; for the non-thermal spectral index, we used
$\alpha_{nth}$=0.7 \citep{haynesetal91}. This simple method of
separation assumes that the emission is optically thin, and that the
synchrotron emission can be described by a single power-law over the
1.4 to 4.8 GHz frequency range. We note that our assumption of a
constant non-thermal spectral index may introduce a bias in the
estimated thermal fraction, since the spectral index is expected to
steepen as cosmic rays propagate from their acceleration sites. If
these acceleration sites are near star-forming regions, then such
regions will have a flatter non-thermal spectral indices than we
assume, and thus the thermal fraction will be
overestimated. Similarly, the thermal fraction will be underestimated
in regions that are far from acceleration sites. However, the uniform
distribution of thermal fractions observed across the sixteen
sub-regions (Fig.~\ref{fig:SRbreakdown}) suggests that this bias is not
severe on the scales that we analyse here. \\

For the radio emission across the whole LMC, we derive a thermal
fraction of 0.45. This increases to 0.57 if we restrict our
field-of-view to the area of the galaxy that belongs to the 16
sub-regions, suggesting that the diffuse radio emission around the
outer edges of the LMC on average has a higher non-thermal fraction
than the radio emission within the galaxy. The difference in flux
density between the 1.4~GHz emission from the 16 sub-regions and the
whole LMC is 62 Jy, so while the radio emission from the galaxy's
edges is intrinsically weak, it still makes some contribution to the
LMC's total flux density (471 Jy). The results of our decomposition
for individual sub-regions are listed in Table~\ref{tbl:SRdata2}. The
thermal fraction varies from $\sim$0.9 for the field containing 30
Doradus, to $\sim$0.1 in the south-west of the LMC (sub-region 2). For
sub-regions 1 and 14, our decomposition produces an unphysical result,
i.e. thermal fractions less than zero. These anomalous values for
$f_{th}$ arise because we fix $\alpha_{nth}$=0.7 in our decomposition;
both fields are dominated by a bright compact radio point source for
which a steeper non-thermal spectral index may be more
appropriate. The median thermal fraction for the remaining sub-regions
is 0.45, which is higher than the typical value for spiral
galaxies \citep[$\sim$0.1,][]{condon92}, but consistent with
radio continuum observations of other dwarf galaxies
\citep{kleinetal91,skillmanklein88}. The sub-regions with the highest
1.4~GHz flux density generally have higher thermal fractions. One exception
is sub-region 4, the field south of 30 Doradus. Here the thermal
fraction is low (0.26), but the 1.4~GHz emission is relatively
strong. Previous studies of the polarisation of the radio emission in
the LMC have shown that the magnetic field is well-ordered in this
region \citep{kleinetal93}. \citet{kleinetal89} have argued that the
higher non-thermal fraction in this part of the LMC arises through a
tidal interaction between the LMC and SMC, which drags the magnetic
field of the LMC towards the SMC and facilitates the diffusion of
relativistic particles into the intergalactic medium.\\

%%%%%%%%%%%%%%%%%%%%%%%%%%%%%%%%%%%%
%%%%%%%%%%%%%%%%%%%%%%%%%%%%%%%%%%%%
%
% TABLE 4
%
%%%%%%%%%%%%%%%%%%%%%%%%%%%%%%%%%%%%%
%%%%%%%%%%%%%%%%%%%%%%%%%%%%%%%%%%%%%

\begin{table}
  \caption{Pearson's correlation coefficient, $r_{p}$, and wavelet
    correlation coefficient, $r_{w}$, at 4 different spatial scales
    between image pairs for the whole LMC. The wavelet scales shown
    are $\log(a)$ = 2.0, 2.5, 3.0 and 3.5, corresponding to spatial
    scales $d=25$, 80, 245 and 770~pc. See also
    Figure~\ref{fig:xcorLMC}.}
  \label{tbl:coefficients} 
\begin{center}
    \begin{tabular}{@{}lrrrrr@{}}
   \hline
   Image pair & $r_{p}$  & $r_{w,2.0}$ & $r_{w,2.5}$ & $r_{w,3.0}$ & $r_{w,3.5}$ \\
     \hline
   60\,$\mu$m-RC & 0.86 & 0.27 & 0.89 & 0.96 & 0.98    \\
   100\,$\mu$m-RC & 0.85 & \nodata & 0.79 & 0.91 & 0.95 \\
   RC-CO & 0.20 & \nodata  & 0.13 & 0.23 & 0.31    \\
   60\,$\mu$m-CO & 0.31 & \nodata & 0.27 & 0.34 & 0.37   \\
   RC-\HI\ & 0.14 & $-0.12$ & $-0.11$ & 0.12 & 0.31   \\
   60\,$\mu$m-\HI\ & 0.31 & 0.06 & 0.03 & 0.22 & 0.33  \\
   100\,$\mu$m-\HI\ & 0.48 & \nodata & 0.16 & 0.35 & 0.45  \\
   CO-\HI\ & 0.37 & \nodata  & 0.12 & 0.40 & 0.56   \\
   \hline
    \end{tabular}
\end{center}
\end{table}

In the bottom panel of Fig.~\ref{fig:SRbreakdown}, we plot the spatial
scale at which the radio-FIR wavelet cross-correlation falls below
0.75 against the thermal fraction of each sub-region. For sub-region
8, where the decorrelation at small scales may be partially due to the
resolution of the HIRAS FIR data, we also plot the result for the
wavelet cross-correlation between the 1.4~GHz and HIRES FIR data (open
circle).  There is a clear inverse relationship between these two
parameters: in sub-regions with a high thermal fraction, the radio-FIR
wavelet cross-correlation holds to much smaller scales.  We also see a
steepening of the WLS fit to the radio-FIR correlation with increasing
thermal fraction (top panel of Fig.~\ref{fig:SRbreakdown}), which is
consistent with the increased FIR/radio ratio near star-forming
regions discussed in Sections~\ref{sect:pixcmp} and \ref{sect:qmap}.\\

%%%%%%%%%%%%%%%%%%%%%%%%%%%%%%%%
%%%%%%%%%%%%%%%%%%%%%%%%%%%%%%%%
%% FIGURE 9                 %%
%%%%%%%%%%%%%%%%%%%%%%%%%%%%%%%%
%%%%%%%%%%%%%%%%%%%%%%%%%%%%%%%%

%%%%%%%%%%%%%%%%%%%%%%%%%%%%%%%%%%%
\section{Discussion}
%%%%%%%%%%%%%%%%%%%%%%%%%%%%%%%%%%%
\label{sect:discussion}

The primary result of this study is that the local radio-FIR
correlation in the LMC is very good: within some regions of the
LMC, the 60~$\mu$m and 1.4~GHz emission remain strongly correlated
down to spatial scales that correspond to the resolution of our data,
i.e. $\sim$20~pc. On almost all scales that we investigate, the radio-FIR
correlation is markedly stronger than the correlation between either
the radio or the FIR emission with the gas surface density. However,
we also find that the best correlation between the radio and FIR
emission occurs in regions where the thermal fraction of the radio
emission is high. These regions typically host the brightest emission
regions in the LMC, where a steeper than linear correlation is also
found. Elsewhere in the LMC, the radio-FIR correlation is flatter, and
tends to show a poorer correlation on small spatial scales. In
this section, we discuss the results of our analysis in relation to
previous studies of the radio-FIR correlation in the LMC. We also
consider our results in the context of recent work on the radio-FIR
correlation in normal spiral galaxies.

\subsection{Comparison with Previous LMC Work}
\label{sect:previous}

The striking morphological resemblance between the FIR and radio
emission has been noted by several previous studies, albeit with lower
resolution data than used in this study
\citep[e.g.][]{kleinetal89,xuetal92}.  In particular, \citet{xuetal92}
have studied the correlation between the FIR and 6.3 cm radio continuum
emission in the LMC using pixel-by-pixel techniques. The limiting
resolution of their data was 4.8 arcmin, allowing them to examine the
correlation on spatial scales above 70~pc (for $d_{{\rm
    LMC}}=50.1$~kpc). Our results are generally in good agreement with
the \citet{xuetal92} results. The values that \citet{xuetal92} derive
for the correlation coefficient ($r_{p}$=0.84) between their FIR and
6.3 cm radio map, and their mean FIR/1.4~GHz ratio ($q_{\rm
  mean}=2.40$) are effectively identical to our measurement of these
quantities ($r_{p}$=0.86, and $q_{\rm median}=2.45$). Likewise, we
find no radial dependence of the FIR/radio ratio. \\

\begin{figure}
\includegraphics[width=85mm]{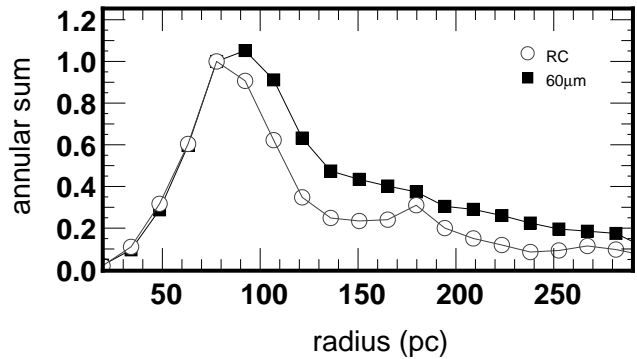}
\caption{Radial profile of the 60~$\mu$m and 1.4~GHz radio emission
  around N11, centred on $\alpha$=4$^{\rm h}$56$^{\rm m}$43$^{\rm s}$,
  $\delta$=$-66$\degr29\arcmin09\arcsec.  The surface brightness in an
  annulus of radius $r$, around the central position for the 60~$\mu$m
  emission (filled squares) and the 1.4~GHz radio emission (open
  circles) is shown. The curves are normalised by their raw value at
  $r \sim 80$~pc, and the width of each annulus is 15~pc.}
\label{fig:N11}
\end{figure}

Through a statistical analysis of 35 \HII\ regions, \citet{xuetal92} have
found that the total 6.3 cm radio emission associated with an
\HII\ region in the LMC is more centrally concentrated than the total
FIR emission, causing the FIR/6.3 cm radio ratio to break down on
scales below 70~pc. The authors argued that the central peak in the
radio emission -- which produces a ``dip'' in the $q$-value -- should
be attributed to thermal radio emission from the ionizing source. At
larger radii ($\sim70-200$~pc), they find the FIR/radio ratio to be
enhanced, corresponding to the characteristic scale of dust heating
around the star-forming region. The \citet[][]{xuetal92} results were
based on a statistical analysis of \HII\ regions in their data, but
our $q$-map clearly confirms their description of a ``dip-and-ring''
morphology for the largest \HII\ regions in the LMC. In
Fig.~\ref{fig:N11}, we present a radial analysis of the radio and
60~$\mu$m emission associated with N11. Both the radio and the
infrared emission show a ring-like morphology: the ring structure is
constituted by both bright, well-known star-forming objects and
diffuse, filamentary emission at $r\sim 50$--100~pc from a centre
position of $\alpha$=4$^{\rm h}$56$^{\rm m}$43,
$\delta$=$-66$\degr29\arcmin09\arcsec (J2000). The 60~$\mu$m emission
is clearly more extended than the radio emission, however, peaking at
larger radii and decaying more gradually towards the background
level. \\

Finally, it is worth noting the discrepancy between our results for
the slope of the local FIR-radio correlation (Fig.~\ref{fig:pixcmpLMC}
and Fig.~\ref{fig:pixcmpSR}) and those of \citet{xuetal92}. An
unweighted least squares fit to the \citet{xuetal92} data indicated a
slope of 0.98$\pm$0.02 for the whole LMC, consistent with a linear
correlation. However, by determining both the WLS fit and the OLS
bisector to our scatterplot of the radio and FIR emission in the whole
LMC, and by examining the emission in different sub-regions across the
galaxy, we identify a change in the slope of the correlation. This
change in slope suggests that there are two distinct physical
mechanisms responsible for the correlation. A steeper relation (slope
$\sim 1.1-1.3$) is found for the high-intensity pixels in the most
active star-forming regions, where the thermal fraction of the radio
emission is relatively high. A flatter relation (slope $\sim 0.6-0.9$)
applies more generally to the diffuse radio and FIR emission across
the LMC, and in regions with a greater non-thermal contribution to the
total radio emission (Fig.~\ref{fig:pixcmpSR} and
Fig.~\ref{fig:SRbreakdown}). The most probable reason for the
discrepancy between our results and those of \citet{xuetal92} is our
ability to sample emission from the LMC more finely, due to the higher
angular resolution of our data. Higher resolution permits a better
distinction between compact and extended emission than would be
possible otherwise. \\

Our results for the slope of the correlation between the radio and FIR
emission are consistent with a cosmic ray diffusion scenario
\citep{bicayhelou90, murphyetal05}, in the sense that while there is
an apparent deficit of local radio emission across the LMC, the LMC's
{\it total} radio flux density is not unusually low. In other words, a
flatter than linear slope for the local correlation is not necessarily
inconsistent with the linear slope of the FIR-radio correlation for
integrated galaxy flux densities, since the integrated flux density
measurements used to determine the global correlation can effectively
recoup the synchrotron emission from the cosmic ray electrons that
have diffused away from their production sites.

%%%%%%%%%%%%%%%%%%%%%%%%%%%%%%%%
%%%%%%%%%%%%%%%%%%%%%%%%%%%%%%%%
%% FIGURE 10         %%
%%%%%%%%%%%%%%%%%%%%%%%%%%%%%%%%
%%%%%%%%%%%%%%%%%%%%%%%%%%%%%%%%

\subsection{Origin of the global radio-FIR correlation}

\subsubsection{Star formation models}

As discussed in \S\ref{sect:intro}, conventional models for the
radio-FIR correlation attribute both sources of luminosity to massive
star formation.  Objections to this explanation are often
based on the fact that the FIR is a prompt and relatively local
measure of the ionising output of massive stars, while the radio
emission results from cosmic ray electron (CRe$^{-}$) energy losses
occurring over a much longer timescale and a much larger area. On a
globally averaged basis and in a steady-state situation, however, a
strong correlation between the radio and FIR emission might still be
expected.  The sensitivity of synchrotron emissivity to magnetic field
strength is not a serious obstacle given that the globally integrated
emission reflects the total energy emitted by an electron in its
radiating lifetime rather than its instantaneous emitted power.  On
{\it local} scales, such models predict that the radio emission should
appear as a smeared version of the FIR emission, aside from
small-scale differences due to temporal fluctuations that violate the
steady-state assumption.  Support for this prediction comes from the
larger radial scale lengths of radio compared to FIR discs
\citep{bicayhelou90,murphyetal05}.\\

Models in this category include the ``calorimeter'' model by
\citet{voelk89}, in which the production of both UV photons
and the relativistic electrons is proportional to the massive star
population. The model assumes that in most galaxies the UV photons
from massive stars are completely absorbed and reprocessed by
interstellar dust, and that relativistic electrons are trapped in the
galaxy and lose all their energy through synchrotron or inverse
Compton scattering.  The energy densities of the magnetic field and
the interstellar radiation field are further assumed to be
proportional, maintaining a constant ratio of synchrotron to inverse
Compton losses.  A subsequent, more general version of the calorimeter
model by \citet{lisenfeldetal96} allows for finite escape
probabilities for relativistic electrons and variable optical depth to
UV photons.\\

With regards to the LMC, three results seem especially relevant.
First, the existence of a strong FIR-radio correlation on local scales
appears to contradict the calorimeter model, although
Fig.~\ref{fig:SRbreakdown} implies that this strong correlation is
mainly driven by the correlation between {\it thermal} radio and FIR
emission, for which a local correlation is naturally expected.
Second, the high thermal fraction of radio emission, coupled with a
higher FIR/radio ratio compared to spiral galaxies, suggests that a
large fraction of relativistic electrons can escape the LMC, violating
the optically thick assumption.  In addition, the LMC should be less
optically thick to UV photons than massive spiral galaxies because of
the lower dust-to-gas ratio \citep{fitzpatrick86} and high porosity
implied by the large number of superbubbles and supershells. \\

The third argument against the calorimeter model in the LMC is the
relatively flat spectral index of $\alpha_{\rm nth}$=0.7, whereas in
the case of efficient cosmic ray trapping one would expect
$\alpha_{\rm nth} \approx 1.1$ \citep{niklasbeck97}.  On the other
hand, the integrated spectral index is a poor diagnostic of CRe$^{-}$
escape if either the contribution of SNRs to the total radio emission or
convective transport of cosmic rays is not negligible
\citep{lisenfeldvoelk00}.  Additional constraints could be provided by
examining the variation in spectral index with distance from
star-forming regions.  Clearly, a more detailed decomposition of the
LMC's radio emission into thermal, diffuse synchrotron, and SNR
emission, using multi-frequency high-resolutions data sets, is
required to fully address this issue.\\

\subsubsection{Magnetic field -- gas density coupling models}

An alternative ``optically thin model'' has been proposed by
\citet{heloubicay93}.  Once again, the dust-heating UV photons and the
cosmic-rays are assumed to have a common origin in massive star
formation, but in this case both photons and cosmic rays can escape
the galaxy. The radio-FIR correlation is enforced by maintaining equal
escape rates for photons and CRe$^{-}$'s, i.e.\ $\tau_0 \propto t_{\rm
  esc}/t_{\rm sync}$, where $\tau_0$ is the UV optical depth, $t_{\rm
  esc}$ is the CRe$^{-}$ escape time scale, and $t_{\rm sync}$ is the
synchrotron loss time scale.  In practice, \citet{heloubicay93}
achieve this by coupling the magnetic field $B$ (important for the
synchrotron emissivity) to the gas density $\rho$ (important for UV
absorption) by the relation $B \propto \rho^{0.5}$, assuming that the
typical distance that a CRe$^{-}$ travels before escape ($l_{\rm
  esc}$) depends only weakly on the disc scale height ($h_0$), $l_{\rm
  esc} \propto h_0^{0.5}$.  Additional assumptions include a nearly
constant dust-to-gas ratio, in agreement with observations
\citep[e.g.,][]{xuhelou96}, and a proportionality between synchrotron
and inverse Compton losses, as in the calorimeter model.  The physical
basis for the $B$-$\rho$ coupling could be amplification of MHD
turbulence until energy equipartition is reached
\citep{grovesetal03}.\\

Other models in this class differ slightly in their prescription for
cosmic-ray confinement (and hence number density).
\citet{heloubicay93} assume a confinement process such that $l_{\rm
  esc} h_0^{-0.5}$ is constant, \citet{niklasbeck97} assume
equipartition between the CRe$^{-}$ density ($n_{CR}$) and magnetic
field, $n_{CR} \propto B^2 \propto \rho$, and \citet{hoernesetal98}
assume a nearly constant CRe$^{-}$ density, as do
\citet{murgiaetal05}.  In spite of these differences, which may simply
reflect our poor understanding of how cosmic rays are distributed, the
basic condition of $B \propto \rho^{0.5}$ is common to all of these
models, and is motivated by the sensitive dependence of synchrotron
emission on the magnetic field strength, bearing in mind the large
range of magnetic field strengths observed in galaxies obeying the
correlation \citep{condonetal91}.  These models have the advantage of
naturally producing a {\it local} radio-FIR correlation, breaking down
only on scales $\sim l_{\rm esc}$ due to cosmic ray diffusion.  The
fundamental $B$-$\rho$ coupling, moreover, could be valid down to the
scales of typical gas clouds ($\sim$100~pc).  Most significantly, the
coupling can even produce a radio-FIR correlation in cases where the
FIR is dominated by cool dust heated by older stellar populations,
although the correlation would not necessarily be linear
\citep{hoernesetal98}.\\

Do our LMC results support the idea of a coupling between magnetic
field and gas density?  Although there is a general morphological
correspondence between the large-scale features of the neutral gas and
the radio and FIR maps, the \HI\ and CO emission are poorly correlated
with the radio emission on scales below $\sim$1~kpc
(Figs.~\ref{fig:xcorLMC}~and~\ref{fig:xcorSR2}).  Even in regions with
significant non-thermal radio fractions, the radio-FIR correlation
cannot be simply derived from a correlation between the \HI\ (which
dominates the total gas content) and the radio continuum.  Note
however that the synchrotron emission depends on the number density of
relativistic electrons as well as the magnetic field strength;
\citet{heloubicay93} constrain $n_{CR}$ by assuming that $l_{\rm esc}
\propto h_0^{0.5}$, but provide only qualitative arguments to justify
this assumption.  If the escape length for cosmic rays in the LMC is
significantly shorter than a nominal value of $\sim2$ kpc -- for
example because of a high porosity of the ISM -- then the galaxy cannot
effectively extract energy from the cosmic rays it generates, even if
$B$ attains values typical of spiral galaxies.  Thus, it is still
possible that $B$-$\rho$ coupling holds in the LMC, but that the
CRe$^{-}$'s are not able to fully explore the dense gas distribution.
Direct measurements of the local magnetic field strength throughout
the LMC, e.g.\ through Faraday rotation \citep{gaensleretal05}, will
be required to rigorously test the $B$-$\rho$ coupling model.\\

We can, however, address the suggestion by \citet{niklasbeck97} that
the FIR emission is coupled to the gas density via the
Kennicutt-Schmidt law, $\rho_{\rm SFR} \propto (\rho_{\rm
  gas})^{1.4}$, and that this combined with $B$-$\rho_{\rm gas}$
coupling leads to a tight radio-FIR correlation.  Our wavelet
cross-correlation analysis shows that the FIR emission is much more
poorly correlated with the gas column density than with the radio
emission, even in sub-regions where the thermal radio fraction is
relatively small.  Thus, it appears unlikely that the
Kennicutt-Schmidt law can serve as a basis for understanding the
radio-FIR correlation.\\

\subsubsection{Connection of FIR-radio correlation to molecular gas}

Recently, \citet{murgiaetal05} have demonstrated that the CO-radio
correlation within spiral galaxies is as strong as the FIR-radio
correlation down to scales of $\sim$100 pc.  They explain both by
arguing that the gas density and magnetic field strength are linked
via equipartition between turbulent and magnetic pressure in a
hydrostatic disc.  However, our results indicate a relatively poor
correlation between CO and 1.4~GHz radio emission within the LMC
(Fig.~\ref{fig:xcorLMC}). We identify several possible reasons for the
difference between our results and those of \citet{murgiaetal05}.
First, the thermal radio fraction $f_{\rm th}$ is much higher for the
LMC than in spiral galaxies, so one expects the CO-radio correlation
on small scales to behave more like the CO-60~$\mu$m correlation,
breaking down near \HII\ regions due to the photodestruction of
molecular clouds.  Second, the fraction of neutral gas mass in
molecular form, $f_{\rm mol}$, is much smaller in the LMC, $\sim$0.05
assuming a Galactic CO-to-H$_2$ conversion factor, compared to
$\ga$0.5 in the central regions of massive galaxies studied by
\citet{murgiaetal05} \citep[cf.][]{wongblitz02}.  Since CO cannot
survive without self-shielding, this results in a much lower volume
filling factor for CO emission, and thus no CO counterpart to the
extended non-thermal radio emission.  Finally, the combination of
strong external pressures exerted on the LMC due to its interaction
with the Milky Way halo and variable internal pressures due to the
expansion of bubbles and supershells suggests that simple estimates of
the hydrostatic pressure will not predict the distribution of CO
clouds.  Indeed, observations of the LMC~4 supershell by
\citet{yamaguchietal01} provide evidence for the formation of
molecular clouds in the swept-up \HI\ surrounding the supershell.\\

%%%%%%%%%%%%%%%%%%%%%%%%%%%%%%%%%%
\section{Conclusions}
%%%%%%%%%%%%%%%%%%%%%%%%%%%%%%%%%%
\label{sect:conclusions}

We have examined the correlation between the 1.4~GHz radio continuum and
60~$\mu$m emission in the LMC, using pixel-by-pixel techniques and a
wavelet cross-correlation method. The correlations between the radio
continuum, FIR emission and the integrated $^{12}$CO and \HI\ emission
have also been studied, in order to assess the relevance of gas density
for the local radio-FIR correlation. The high angular
resolution datasets that are available for the LMC allow us to probe
the radio-FIR correlation on spatial scales above $\sim20$~pc, and to
identify variations of the FIR/radio ratio for different interstellar
environments. \\

The main results of our study are:\\

1. Pearson's correlation coefficient shows that the 1.4~GHz and
60/100$\mu$m are positively correlated across the LMC. The correlation
coefficient for 1.4~GHz and 60$\mu$m emission in the whole LMC is
$\sim0.86$, indicating a strong correlation. The correlation improves
in regions where there are clear signs of star-forming activity,
e.g. near 30 Doradus. The correlation between the radio and FIR
emission is significantly stronger than correlations between
these wavebands and cold gas tracers. \\

2. Pixel-by-pixel scatterplots indicate that locally there are two
independent correlations between the radio and FIR emission, both of
which are non-linear. A steeper than linear correlation (slope $\sim
1.1-1.3$) is found for the high-intensity pixels in the most active
star-forming regions, where the thermal fraction of the radio emission
is relatively high. A flatter correlation (slope $\sim 0.6-0.9$)
applies more generally to the diffuse radio and FIR emission across
the LMC, and in regions with a greater non-thermal contribution to the
total radio emission.\\

3.  Both the integrated FIR/radio ratio ($q=2.55$) and the median
FIR/radio ratio across the LMC ($q_{\rm median}=2.45$) are higher than
the mean FIR/radio ratio of the \citet{yunetal01} galaxy sample
($q_{\rm mean}=2.34$), suggesting that the radio emission in
the LMC is slightly underluminous, or that there is a slight excess of
FIR emission. In contrast to nearby spiral galaxies, the brightest
structures in the FIR/radio ratio map of the LMC do not correspond to
the brightest structures in the input FIR and radio maps. \\

4. A simple decomposition of the 1.4~GHz radio emission indicates that
the thermal fraction of the LMC is higher than for normal disc
galaxies \citep[$\sim0.1$,][]{condon92}. For an assumed non-thermal
spectral index of $\alpha_{nth}$=0.7, the thermal fraction for the
whole the LMC is 0.45, varying between 0 and 0.9 for individual
sub-regions. The high FIR/radio ratio and high thermal fraction of the
LMC's radio emission suggests that a large proportion of CRe$^{-}$s
may be escaping from the LMC with low synchrotron losses.\\

5. Wavelet cross-correlation spectra show that the correlation between
the 1.4~GHz radio continuum and 60~$\mu$m emission is very good across
the LMC, even on scales corresponding to a few tens of parsecs. For
the whole LMC, the correlation between the radio and 60~$\mu$m
emission on all scales is significantly better than the correlation of
the \HI\ or CO with either the radio or FIR emission. We argue,
however, that the excellent correlation between the radio and FIR
emission for the LMC as a whole is driven by the 30 Doradus
region. For individual sub-regions within the LMC, there is an inverse
relationship between the scale on which the correlation breaks down
and the thermal fraction of the radio emission: regions that show a
strong correlation to small spatial scales are also the regions where
the thermal fraction of the radio emission is high. In regions where
the thermal fraction is less than 0.5, the wavelet cross-correlation
between radio and 60~$\mu$m emission breaks down across a range of
scales between 50 and 250\,pc. \\

While models for the radio-FIR correlation typically address the
emission and coupling mechanisms in normal spiral galaxies, the LMC
remains an interesting testbed because of the wide range of angular
scales that can be probed.  The presence of a strong local radio-FIR
correlation that is dominated by high-brightness thermal emission,
combined with evidence for a diffuse non-thermal radio component that
contributes to the LMC's total radio flux density, underscores that
different phenomena can be responsible for brightness as opposed to
integrated flux density.  This distinction should be kept in mind in future
studies of more massive galaxies.\\

%%%%%%%%%%%%%%%%%%%%%%
%
% ACKNOWLEDGEMENTS
%
%%%%%%%%%%%%%%%%%%%%%%%

\label{sect: thanks}
{\it We would like to thank the referee, John Dickel, for his careful
  reading and astute comments that improved the final version of this
  paper significantly. We also thank Jason Surace, Michael Braun,
  Sunguen Kim, John Dickel, Uli Klein, and Marc-Antoine
  Miville-Desch\^{e}nes for access to the datasets that were used in this
  study. Rainer Beck and Eric Murphy contacted us about this work
  prior to publication, and our understanding has benefitted from their
  insight and expertise. AH enjoyed illuminating discussions with
  Vince McIntyre, Steve Ord, Juergen Ott, Erik Muller, Jean-Phillipe
  Bernard, Roberta Paladini and Michael Dahlem, and would like to
  thank Christophe Pichon and Eric Thi\'{e}baut for developing many
  invaluable \textsc{Yorick} routines that were used in this work.}

\label{sect:bibliography}
\bibliographystyle{mn}
\bibliography{radiofir,lmc,wavelets,obstools,mnemonic,software}

\label{lastpage}

\end{document}